\documentclass[useAMS,usenatbib]{mn2e}
\usepackage{amssymb}
\usepackage{amsmath}
\usepackage{cases}
\usepackage{graphicx}
\usepackage{lscape}
\usepackage{color}
\usepackage{subfigure}

\def\h2{\rm{H_2}}
\def\fh2{f_{\rm{H_2}}}

\def\Sh2{\Sigma_{\h2}}

\def\ms{M_{\odot}}
\def\tti{t_{\rm TI}}
\def\tff{t_{\rm ff}}
\def\LX{L_{\rm X}}
\def\TX{T_{\rm X}}
\def\Mvir{M_{\rm 200}}

\newcommand{\eg}[0]{$\textnormal{e.g. }$}
\newcommand{\ie}[0]{$\textnormal{i.e. }$}

\title[Hot gas models in SAMs]{The hot gas distribution, X-ray luminosity and baryon budget in the L-Galaxies semi-analytic model of galaxy formation}

\author[Zhong et al.]{Wenxin Zhong$^{1,2}$, Jian Fu$^{1}$\thanks{E-mail: fujian@shao.ac.cn}, Prateek Sharma$^3$, Shiyin Shen$^{1,4}$,  Robert M. Yates$^{5,6}$\\
$^1$Key Laboratory for Research in Galaxies and Cosmology, Shanghai Astronomical Observatory, CAS,\\ 80 Nandan Rd., Shanghai, 200030, China PR\\
$^2$University of Chinese Academy of Sciences, No. 19A Yuquan Road, 100049, Beijing, China PR\\
$^3$Department of Physics and Joint Astronomy Program, Indian Institute of Science, 560012, Bangalore, India\\
$^4$Key Lab for Astrophysics, Shanghai, 200034, Shanghai, China PR\\
$^5$Astrophysics Research Group, University of Surrey, Stag Hill, Guildford, GU2 7XH, UK\\
$^6$Centre for Astrophysics Research, University of Hertfordshire, Hatfield, AL10 9AB, UK
}

\begin{document}

\maketitle

\begin{abstract}

Hot ionized gas is important in the baryon cycle of galaxies and contributes the majority of their ``missing baryons''. Until now, most semi-analytic models of galaxy formation have paid little attention to hot gaseous haloes and their X-ray emission. In this paper, we adopt the one-dimensional model from Sharma et al. instead of the isothermal sphere to describe the radial distribution of hot gas in the L-Galaxies semi-analytic model. The hot gas halo can be divided into two parts according to the ratio of the local thermal instability time-scale and the free-fall time-scale: a cool core with $\tti/\tff=10$ and a stable outer halo with $\tti/\tff>10$. We update the prescriptions of cooling, feedback and stripping based on the new hot gas profiles, and then reproduce several X-ray observational results, like the radial profiles of hot gas density, and the scaling relations of X-ray luminosity and temperature. We find: (1) Consistent with observations, flatter density profiles in halo centers produce lower X-ray emission than an isothermal sphere; (2) Cool core regions prone to precipitation have higher gas temperature than the virial temperature, and a larger $\TX/T_{\rm 200}$ ratio in smaller haloes leads to a steeper slope in the $\LX-\TX$ relation; (3) The ionized gas in the unbounded reservoir and low temperature intergalactic gas in low mass haloes could be the main components of the halo ``missing baryons''. Our model outputs can predict the observations of hot gas in the nearby universe and produce mock surveys of baryons probed by future X-ray telescopes.

\end{abstract}



\section{Introduction} \label{sec:intro}

According to the current picture of cosmology, the hot ionized gas contributes most of the baryon budget in the universe. The reservoir of hot gas in the circumgalactic and intracluster medium (ICM and CGM) plays an important role in galaxy baryon cycles like gas accretion, star formation, and feedback (see review paper by Tumlinson et al. 2017). 
Based on $\Lambda$CDM cosmology, the universal baryon density fraction is around 4.8\% (Planck collaboration et al. 2016), while the observed baryons in galaxies (stars, interstellar medium (ISM)), cluster environments (CGM, ICM), and Lyman-$\alpha$ absorbers only contribute about 60\% of the total baryon budget, and around 30-40\% of the baryons are ``missing''. This is the ``missing baryon problem'' (Shull, Smith \& Danforth 2012). Some works (e. g.,  Martizzi et al. 2019, Kovacs et al. 2019) suggest that the missing baryons are mainly located as diffuse ionized gas in the warm-hot intergalactic medium (WHIM) with temperature $10^5-10^7$K, which is very difficult to observe. In addition to this global missing baryon problem, individual haloes like the Milky Way also seem to be missing their baryons in diffuse phases that are hard to detect. This is the ``local missing baryon problem" (Bregman 2007).

To trace ionized gas components around galaxies and clusters, the observations of X-ray emission/absorption or UV absorption are usually adopted. For ICM gas in large haloes ($\Mvir>10^{12}\ms$), the hot gas with $T>10^6$K tends to emit radiation in the soft X-ray band (Gursky 1971), and the temperature, luminosity, and mass of hot gas can be estimated through X-ray emission. In recent years, X-ray telescopes with high sensitivity and resolution like \emph{XMM-Newton} and \emph{Chandra} have obtained CGM and ICM emission for a large sample of galaxies (Bogd\'{a}n et al. 2015; Li \& Wang 2013; Li et al. 2017; Babyk et al. 2018). However, the gas between $10^4$K and $10^6$K, which is usually called the cool or warm phase, does not emit X-rays. This cool/warm gas is typically probed through UV absorption-lines with background quasars. Powerful UV spectrographs like COS on the Hubble Space Telescope (\eg{}Tumlinson et al. 2013; Fang et al. 2014; Richter et al. 2017; Zahedy et al. 2019) provide a lot of information on various properties of the gaseous absorbers, including the column density, metal abundance, spatial distribution, etc.

Numerical simulations and semi-analytical models offer alternative methods to study the evolution of galaxies. In recent years, large-scale cosmological hydrodynamical simulations in Mpc-scale boxes offer a way to study the physics of galaxy formation. Cosmological galaxy formation simulation projects like EAGLE (Crain et al. 2015; Schaye et al. 2015) and Illustris-TNG (Nelson et al. 2018a; Springel et al. 2018) offer precise simulated statistical samples to study the physical process in the ICM and CGM (e. g., Stevens et al. 2017, Kelly et al. 2021, Nelson et al. 2018b; Truong, Pillepich \& Werner 2021). Because of the high resolution and detailed subgrid physics of gas processes, very high computational resources are needed to run hydrodynamical simulations in a cosmological volume. It is difficult to analyse the effect of physical processes and model parameters on specific properties of the galaxies in these simulations (see review paper by Somerville et al. 2015). Over the past decades, semi-analytic models (hereafter SAMs) have been used as flexible and low-cost tools to study galaxy formation processes, which are usually described by equations from empirical fitting or ``phenomenological'' models. By considering and linking physical processes such as mergers, gas cooling, accretion, star formation, feedback, stripping, and others, SAMs have achieved great success in recent years [\eg{}L-Galaxies 2015 (Henriques et al. 2015); L-Galaxies 2020 (Henriques et al. 2020); Shark (Lagos et al. 2018); SAGE (Stevens, Croton \& Mutch 2016)]. However, most of these models pay little attention to the distribution and observational properties of the ionized gas phase and therefore give poor predictions for the X-ray emission from the CGM surrounding galaxies.

In SAMs, the ionized gas is usually regarded simply as ``hot gas'' within the dark matter haloes around galaxies. Most SAMs assume that hot gas is distributed in a singular isothermal sphere, \eg{}L-Galaxies (De Lucia \& Blaizot 2007, Guo et al. 2011, Henriques et al. 2015, Henriques et al. 2020), GAEA (Hirschmann et al. 2016), the Santa Cruz model (Somerville et al. 2008), Shark (Lagos et al. 2018), and SAGE (Croton et al. 2016). For the isothermal sphere, the density profile is written as
\begin{equation} \label{eq:isothermal}
{\rho _{{\rm{hot}}}}\left( r \right) = \frac{m_{{\rm{hot}}}}{4\pi {r_{{\rm{\rm 200}}}}}{r^{-2}},
\end{equation}
where $m_{\rm hot}$ and $r_{\rm 200}$ are the mass of hot gas and virial radius of the galaxy halo. In such an oversimplified model, the temperature of the hot gas is assumed to be equal to the virial temperature of the halo (White \& Frenk 1991). Based on more detailed theoretical modelling and observations of the hot gas surrounding galaxy groups and clusters (Mulchaey 2000), a more realistic $\beta$ profile is often used to model the hot gas profile, \eg{}in the GALFORM model (Font et al. 2008) and GALACTICUS (Benson 2012), which is given by
\begin{equation} \label{eq:betaprofile}
 {\rho _{{\rm{hot}}}}\left( r \right) = \frac{{{\rho _0}}}{{{{\left[ {1 + {{\left( {r/{r_{{\rm{core}}}}} \right)}^2}} \right]}^{3\beta /2}}}},
\end{equation}
where $\rho_0$ is the central density and $r_{\rm core}$ is the core radius. The $\beta$-profile models are based on the assumption that the hot gas is isothermal and in hydrostatic equilibrium (Cavaliere \& Fusco-Femiano 1976). The MORGANA model by Monaco, Fontanot \& Taffoni (2007) adopts a polytropic equation of state in an NFW halo, and predicts the radial distribution of hot gas, pressure and temperature. However, they did not perform a detailed analysis and comparison of the results of this complex hot gas distribution model with  X-ray observations. Such a $\beta$-profile model was also adopted by Yates et al. (2017) in their study of galaxy groups and clusters using the L-Galaxies SAM, but only as a post-processing step to calculate mean ICM temperatures and iron abundances.

In this paper, we develop a new extension of SAMs to study the hot ionized gas in haloes. Our model is based on the Munich L-Galaxies framework described in Henriques et al. (2015, hereafter H15), which runs on the halo merger trees of the Millennium (Springel et al. 2005) and Millennium II (Boylan-Kolchin et al. 2009) simulations. Our main update is on the radial distribution of the hot gas component in each dark matter halo. We adopt a realistic recipe by Sharma et al. (2012b, hereafter Sharma12), which considers hydrostatic equilibrium and rough global thermal balance in cool cores, and predicts the existence of a low density core in the central region, especially for low-mass haloes. The Sharma12 prescription for the state of hot gas in halos is based on simulations of atmospheres in global thermal balance (motivated by a lack of cooling flows in galaxy clusters and the presence of X-ray cavities that can roughly compensate for radiative losses), which show that cold gas condenses out of hot hydrostatic atmospheres if the ratio of the thermal instability time (similar to the cooling time for free-free emission) and the free-fall time ($\tti/\tff$) is smaller than a threshold close to 10 (Sharma et al. 2012a). The cold gas produced in the core can be accreted by the central supermassive black hole and can act as a valve that maintains the cool core in a ($\tti/\tff \approx 10$) state. More realistic AGN jet feedback simulations (\eg{}Prasad et al. 2015, Li et al. 2015, Ehlert et al. 2022) are roughly consistent with the Sharma12 models in that the $\tti/\tff$ ratio stays within a factor of a few from the fiducial value of 10. These models are also known as precipitation models and have been applied to the CGM of Milky Way mass halos (Voit et al. 2019).

In our new model, we also adjust the prescriptions of the baryonic processes related to the hot gas like feedback, gas cooling and stripping based on the new radial profiles, and output the radial distributions of the hot gas density, temperature and X-ray luminosity in each halo so that detailed comparisons can be made with the latest observations.

The main motivation of the paper is to include more realistic models of hot gas components in SAMs, and offer detailed predictions for the properties of halo hot gas in the X-ray band. This paper should help us compare the hot gas properties, like the gas radial profiles and scaling relations between gas temperature and X-ray luminosity. In addition, we will also study the missing baryon problem.

This paper is organized as follows. In Section 2, we briefly describe the L-Galaxies model, how we handle the radial distribution of halo hot gas, and the corresponding changes to the physical recipes. In Section 3, we show the model results of the radial density profiles, X-ray luminosity and temperature of the hot gas compared with observations, and analyse the physical mechanisms related to the hot gas. In Section 4, we analyse the baryon budget in different halo components, which can help us study the missing baryon problem in the model. In Section 5, we summarize our results and look forward to future work.

\section{The galaxy formation models} \label{sec:model}

\begin{figure}
\centering
 \includegraphics[angle=0,scale=0.4]{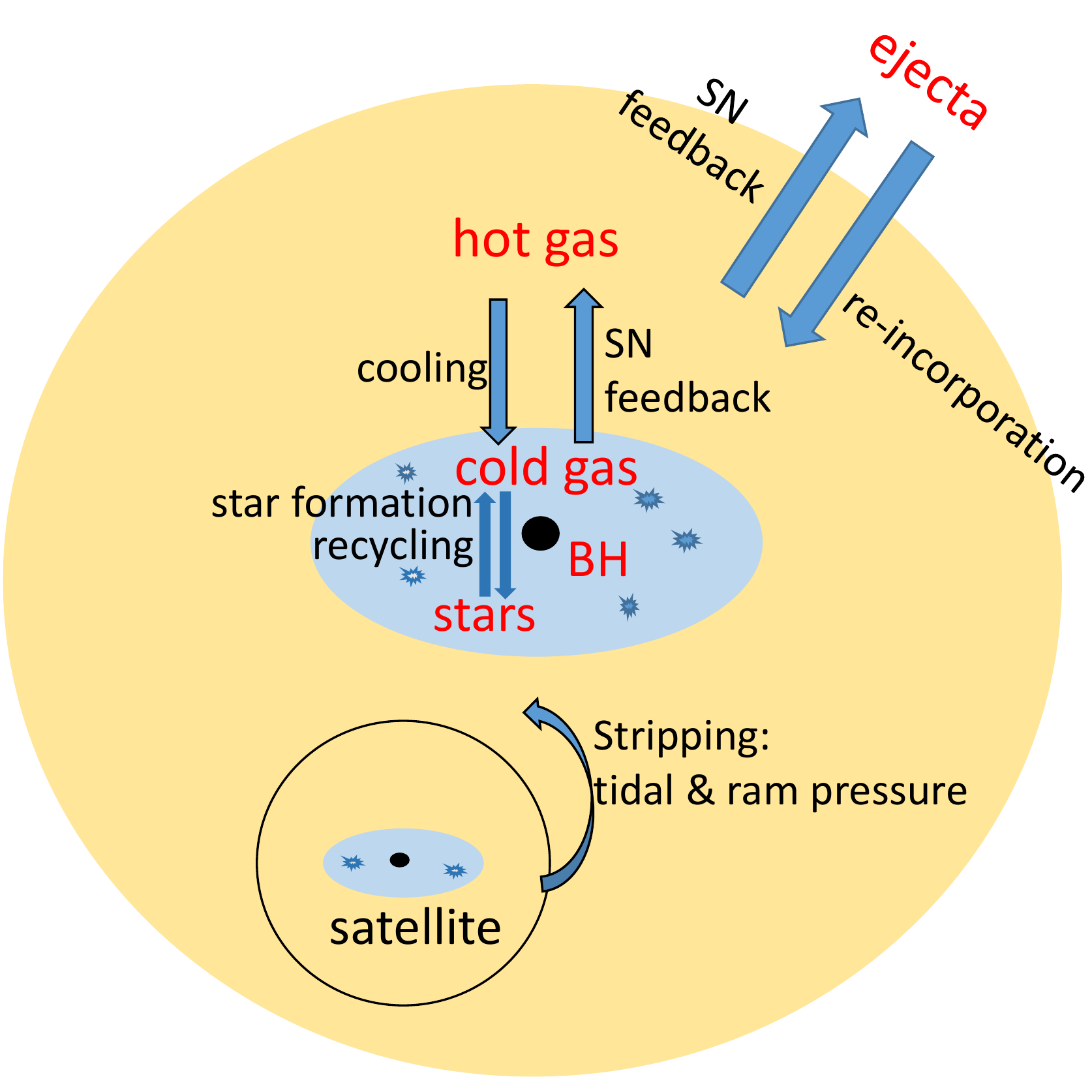}
 \caption{Baryonic processes in L-Galaxies models. The yellow area represents the hot gas in the halo and the blue area represents the cold gas in ISM of a galaxy disk. The red texts show the baryon components of a galaxy or cluster, and the arrows show the exchange and transition processes among these components.}\label{fig:procceses}
\end{figure}

This study is based on the H15 version of the L-Galaxies SAM of galaxy formation. The reader can find the full descriptions of the astrophysics processes in H15. The model codes,  written in C-language,  are publicly available on the L-Galaxies webpage\footnote{https://lgalaxiespublicrelease.github.io}. In Fig. \ref{fig:procceses}, we show a brief illustration of the baryonic processes in L-Galaxies, especially the processes related to the hot gas components. In each galaxy, the baryons are distributed into the following components: disc stars, bulge stars, a central supermassive black hole, cold gas in ISM, ionized gas in a hot gaseous halo, and an external reservoir (ejecta) beyond the halo potential. The arrows in Fig. \ref{fig:procceses} show the exchange of mass and energy among these components.

When a dark matter halo is first identified in the underlying Millennium or Millennium II simulation, it is seeded with hot gas of primordial metallicity. Due to radiative cooling, part of the hot gas loses pressure support and falls to the centre of the halo. The infalling gas cools and can form stars. We also assume the ``instantaneous recycling approximation'', whereby 43\% of the mass formed into stars is immediately returned to the gas phase as a mix of unprocessed material (95.4\%) and newly-synthesised metals (4.6\%). 
Supernova can reheat part of the cold gas to the temperature of the hot gaseous halo, and also eject part of the hot gas out of the halo if the supernova energy is large enough. The ejected mass forms an external reservoir and may fall back into the halo again at a later time. The black hole is embedded in the centre of a galaxy, and the hot gas accreted by the central black hole can power the active galactic nucleus (AGN), which also regulates the cooling onto the galaxy. Satellite galaxies can be influenced by various environmental effects in the halo (ram pressure stripping by ICM and tidal disruption by the gravity of the central galaxy), and may merge with the central galaxy through dynamical friction. When two galaxies merge, a merger-induced starburst will happen and consume a large proportion of the ISM. A major merger in L-Galaxies will also destroy the disk structure of the central galaxy and form a bulge dominant galaxy. These physical processes are implemented as a set of differential equations, and we calculate these physical processes and update the results of the baryonic components in each time step (the time step is around 10-20~Myr, which is about the life time of high mass stars). In Sec. \ref{sec:hotprofilemodel} to \ref{sec:sfmodel}, we describe the prescriptions of the processes related to the hot gas component, and the details of other processes can be found in the H15 paper or the L-Galaxies webpage.

The model is based on the dark matter haloes from N-body simulations. Similar to previous L-Galaxies model papers (\eg{}Guo et al. 2013; Fu et al. 2013; Henriques et al. 15; Henriques et al. 2020), we adopt the dark matter haloes from the Millennium Simulation (hereafter MS, Springel et al. 2005) and Millennium II Simulation (hereafter MS-II, Boylan-Kolchin et al. 2009). The original MS and MS-II adopt the cosmology parameters from WMAP1 (Spergel et al. 2003) and we rescale to the $\Lambda$CDM cosmology with parameters from Planck ($\Omega_\Lambda=0.685, ~\Omega_m=0.315,~\Omega_{\rm{baryon}}=0.0487, ~\sigma_8=0.829$ and $h=0.673$, Planck Collaboration 2016) following the techniques developed by Angulo \& Hilbert (2015). The box size in MS is about 480 Mpc$/h$ (when rescaled) on a side and the minimum halo mass is about $2.9\times10^{10}\ms$ (the mass of 20 particles) and MS-II has a smaller ($\sim{}96$ Mpc$/h$) simulated box and 125 times higher resolution compared to MS ($2.3\times10^8\ms$ for minimum halo mass). In our work, we adopt MS for galaxies with stellar mass $M_{*}>10^{9}\ms$ and MS-II for smaller halos.

In our new model, we incorporate the radial distribution of hot gas in halos by Sharma et al. (2012b) into L-Galaxies. Since the model only contains a cold gas component in galaxies and a hot gas component in the halo, we will not distinguish between the ionized hot gas components in the ICM of the main halo and the CGM of the satellites, but treat both as ``halo hot gas'' in the model. \footnote{SAMs do not consider the details of structures like filaments, knots and WHIM. These structures in the model are thought to reside in hot gas halo or the ejecta reservoir depending on whether or not they are bounded within the halo potential.}

Our main changes are: (1) We abandon the isothermal distribution approximation for the hot gaseous halo, and adopt the one-dimensional model of Sharma12, considering hydrostatic equilibrium and self-regulation by thermal instability in the cool core; (2) We modify the gas cooling and infall prescription based on the new model of the hot gas halo, which is related to the existence of a ``cool core'' in inner halo; (3) We update the prescriptions of gas disruption and feedback based on the new radial profiles of hot gas; (4) We include a simple model of atomic-to-molecular gas transition in the ISM and relate the star formation to the mass of molecular gas. We will describe these changes in detail below.

\subsection{Modelling the radial distribution of hot gas in the haloes} \label{sec:hotprofilemodel}

In the previous version of the L-Galaxies SAMs (\eg{}De Lucia \& Blaizot. 2007, Guo et al. 2011, Henriques et al. 2015; Henriques et al. 2020), the ionized hot gas component is distributed as an isothermal sphere in each dark matter halo with $\rho_{\rm hot}\propto r^{-2}$, and the gas temperature equals the virial temperature of the halo $T_{\rm hot}=T_{\rm 200}$.\footnote{The subscript ``200'' here and the following paper represents the values for the halo. $r_{\rm 200}$ is the radius within which the average density of a halo is 200 times the cosmic critical density, and $m_{\rm 200}$ is the mass within $r_{\rm 200}$. In L-Galaxies models, we assume a flat rotation curve for each halo, and the circular velocity is defined as $v_{\rm 200}$. $T_{200}=35.9\left(v_{200}/\rm{km~s}^{-1}\right)^2$K is the virial temperature inside $r_{200}$.} In the new model, we apply the one-dimensional models of Sharma12 to describe the radial profiles of the halo hot gas.

In Sharma12, they assume the hot gas distributes spherically and construct one-dimensional hot gas profiles for the ICM in dark matter haloes. The model assumes that heating and cooling roughly balance globally in the cool core and regulate the core to thermal equilibrium without investigating the physics of how the hot gas is heated. Considering the local thermal instability time-scale $\tti$ and free-fall time $\tff$, the physics of local thermal instability in stratified atmospheres gives a rough upper limit on the density of hot gas which satisfies $t_{\rm TI}/t_{\rm ff}\gtrsim10$ everywhere.

In their model, they first assume an ``unmodified'' gas profile of a generalized NFW form
\begin{equation} \label{eq:genNFW}
\rho=\frac{{{N_{\rm{c}}}}}{{r/{r_{\rm{s}}}{{\left( {1 + r/{r_{\rm{s}}}} \right)}^{-s-1}}}},
\end{equation}
where $N_{\rm{c}}$ is a normalization constant and $s$ is the asymptotic gas density slope $s=d\ln \rho /d\ln r$. According to the ratio $t_{\rm TI}/t_{\rm ff}$ at each radius, the hot gas halo can be divided into two parts: the cool core and the stable region. In the cool core region, the ratio $t_{\rm TI}/t_{\rm ff}$ is smaller than 10 and the local thermal instability becomes nonlinear and produces multiphase gas. Cold filaments condense out of the hot phase and an overdense blob cools and falls to the halo centre. This cooling picture is similar to the two-phase models by Maller \& Bullock (2004). In the stable region where the ratio $t_{\rm TI}/t_{\rm ff}$ is larger than 10, no multiphase gas is expected and the gas is stable without condensation and infall.

For the haloes with cool cores, they tried two prescriptions to ``modify'' the gas profiles. One is an isentropic core with an entropy floor ($T/\rho^{2/3}=$constant); the other is a core with fixed $\tti/\tff=10$, which agrees with the observations of cluster cool cores. The outer boundary condition in each halo is determined by the pressure at $r_{200}$. For haloes with $\tti/\tff>10$ at $r_{200}$, the boundary temperature is $T_{\rm 200}$ and the outer pressure is specified as $p_{200}=k_{\rm s}\langle n \rangle k_{\rm B}T_{\rm 200}$, where $\langle n \rangle$ is the average gas density within $r_{200}$, and $k_{\rm s}$ is a constant. In low mass haloes ($M_{\rm halo}\lesssim10^{13}\ms$), the ratio $\tti/\tff$ is smaller than 10 for the whole halo if we still adopt $T_{200}$ at the halo boundary. In Sharma et al. (2012a), they argue that the multiphase gas will lead to very high accretion rates if $\tti/\tff<10$, and the cool core should self-regulate to the critical value of $\tti/\tff\sim10$ by cooling and feedback. In this case, the models adjust the density and temperature (holding a constant pressure) to modify the profiles such that $\tti/\tff$ equals 10 at the outer boundary (virial radius).

\begin{figure*}
\centering
 \includegraphics[angle=0,scale=0.4]{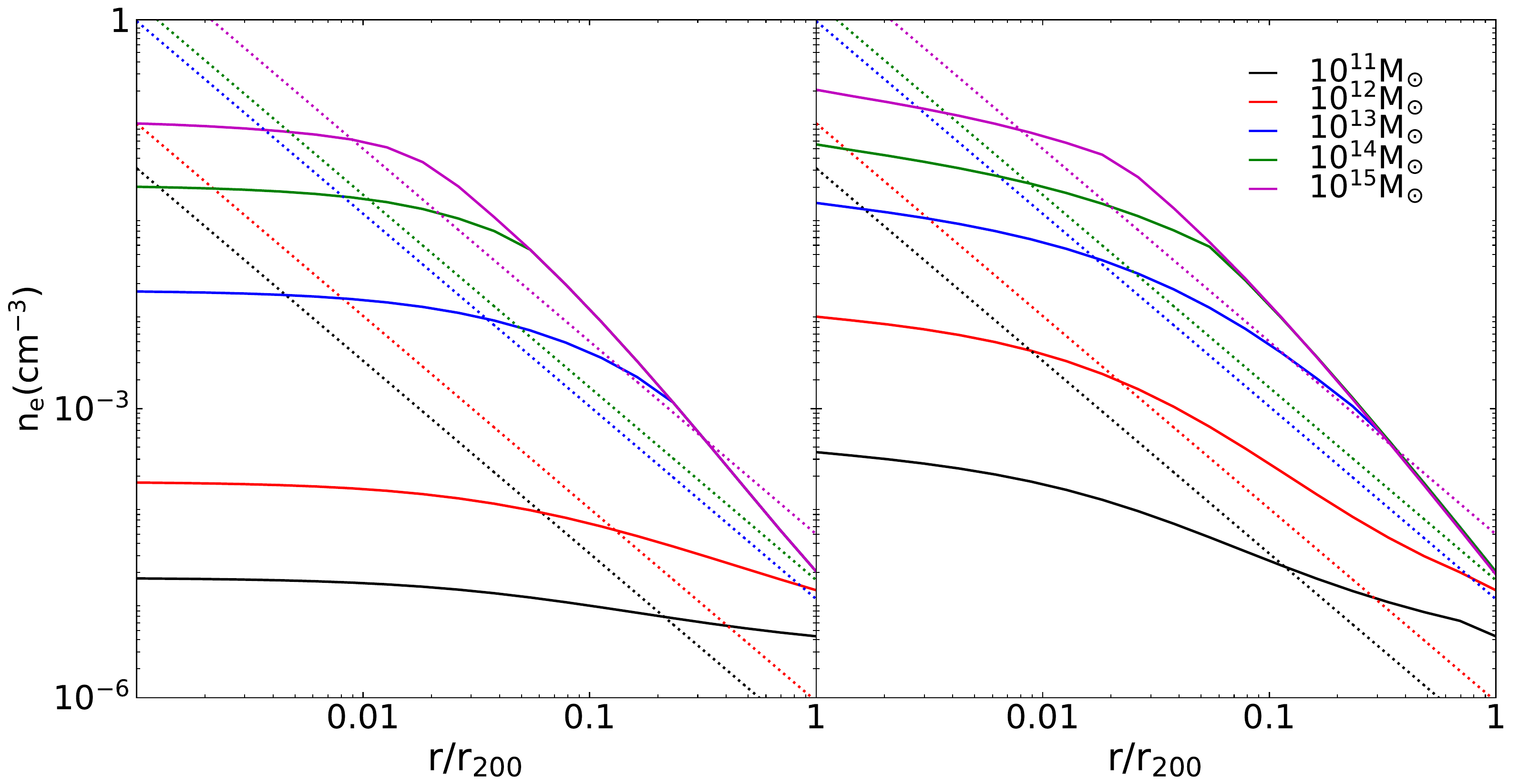}
 \caption{Radial profiles of the electron density from Sharma12 models (solid curves) and the isothermal spheres (dotted curves). The left and right panels show the model profiles with isentropic core (left) and ``$\tti/\tff=10$'' core (right), respectively. In each panel, the curves in different colors represent the results from different halo masses. For model haloes with $M_{\rm halo}=10^{11}, 10^{12}, 10^{13}, 10^{14}, 10^{15}\ms$, we adopt $c_{200}=39, 36, 33, 30, 27$, and we assume $f_{\rm b}=0.17$ and $f_{\rm d}=0$ as the model parameters in the figures.}\label{fig:neprofiles}
\end{figure*}

In Fig. \ref{fig:neprofiles}, we compare the radial profiles of the electron density from Sharma12 models with the isothermal profiles. The left and right panels show the model profiles with isentropic core and ``$\tti/\tff=10$'' core, respectively. We can see that the radial profiles from Sharma12 model in inner haloes are much flatter than the isothermal profile, and are consistent with the profiles in haloes over $10^{13}\ms$ from X-ray observations (see Figure 2 in Sharma12 and our model results in Fig. \ref{fig:netempprofiles} in Sec. \ref{sec:profilesresult}). In haloes with $M_{\rm halo}\sim10^{12}\ms$ or lower, the radial profiles of the two cool-core models show large differences, and the isentropic core model predicts very flat cool-core profiles. In Sharma12, they mainly focus on the results in massive haloes, while our SAMs are applied to a very wide mass range. We will show some comparisons between two cool core models for the scaling relation of the X-ray luminosity in Sec. \ref{sec:Lx}, and find better results with a $\tti/\tff=10$ core than an isentropic core in low mass haloes (see the discussion in Fig. \ref{fig:Lxhalorelation}). In the following, we will adopt the prescription of cool core with $\tti/\tff=10$ as our ``fiducial'' model.

In Sharma12 models, the asymptotic gas density slope $s$ in Eq. \ref{eq:genNFW} is a free parameter, and they choose $s=-1.75, -2.25, -3$. In our current work, we choose $s=-3$, and find that this choice together with a $\tti/\tff=10$ core can give good fits for the X-ray scaling relations (see Fig. \ref{fig:Lxhalorelation} in Sec. \ref{sec:Lx}). On the other hand, the ``unmodified'' gas profile with $s=-3$ is the standard NFW profiles in Eq. \ref{eq:genNFW}, and the hot gas profile at large radii is of the form $\rho\propto\left({r/{r_{\rm{s}}}}\right)^{-1}{\left( {1 + r/{r_{\rm{s}}}} \right)^{ - 2}}$.

Sharma12 models only consider the processes of thermodynamic and hydrostatic physics in hot gas, but do not consider  baryon cycles related to the galaxy evolution like gas accretion, star formation supernova feedback and AGN feedback. The models adopt some input parameters estimated from observational results, which are the halo mass $M_{200}$, the baryon fraction $f_{\rm b}$, 
and the concentration parameter $c=r_{\rm 200}/r_{\rm s}$ ($r_{\rm s}$ is the scale radius). The model outputs are the radial profiles of electron density $n_e(r)$ and temperature $T(r)$ of hot gas at different redshift.

Sharma12 models are written in Matlab or Python codes\footnote{https://bitbucket.org/prats7up/hot\_halos\_2012/src/master/}, which can be easily incorporated in the L-Galaxies model codes. To incorporate Sharma12 model into the L-Galaxies recipes, first we need to ``construct'' the radial distribution of the hot gas components around each galaxies. Following the methods in the previous work in L-Galaxies (e. g., Fu et al. 2009, 2010, 2013, Henriques et al. 2020), we divide each hot gaseous halo into a set of radial concentric shells around the center, and the outer radius $r_{i}$ of each shell is given by the geometric series
\begin{equation} \label{eq:radial}
r_{i}=[1.438^{i-1}\times10^{-3}]{r_{\rm 200}}~~ (i=1,2...20),
\end{equation}
The innermost shell in Eq. \ref{eq:radial} has a radius around $10^{-3}r_{\rm 200}$, while the radius of outermost shell is $r_{\rm 200}$. We have tested other form of the shell division and find the results are quite insensitive to the sub-division of the halo if the shell number is large enough ($i\gtrsim10$).

Then, we combine the codes of Sharma12 into L-Galaxies codes. To get the input values $M_{200}$, $f_{\rm b}$, $c$, $f_{\rm d}$ 
in Sharma12 model, we adopt the physical quantities from SAMs in each time step: 
\begin{equation}\label{eq:input}
\begin{array}{l}
M_{\rm{200}} = {m_{{\rm{200}}}}\\
f_{\rm{b}} = {m_{{\rm{baryon}}}}/{m_{{\rm{200}}}}\\
c = {r_{\rm 200}}/{r_{\rm s}} = \sqrt 2 {\lambda ^{-1}}\\
f_{\rm{d}} = 1 - {m_{{\rm{hot}}}}/m_{\rm{baryon}},\\
\end{array}
\end{equation}
where $m_{\rm 200}$ is the virial mass of the central halo, $m_{\rm hot}$ is the hot gas mass around each galaxy, and the total baryon mass $m_{\rm baryon}$ is the sum of stellar mass, cold gas mass in ISM, black hole mass, and hot gas mass in the halo. In Eq. \ref{eq:input}, ${r_{\rm{s}}} = \lambda{r_{\rm{200}}}/\sqrt 2$ is the scale length of a galaxy according to Mo, Mao \& White (1998), where the spin parameter $\lambda$ is defined as (Bullock et al. 2001) 
\begin{equation}\label{eq:lambda}
\lambda  = \frac{J}{{\sqrt 2 {m_{\rm 200}}{r_{\rm 200}}{v_{{\rm{200}}}}}}.
\end{equation}
The angular momentum $J$, $m_{\rm 200}$, $r_{\rm 200}$ and $v_{\rm 200}$ are directly from dark matter haloes of the N-body simulations.

With the model parameters in Eq. \ref{eq:input} and radial profiles of hot gas in each halo, we can get the radial profiles of the electron density $n_e(r)$ and gas temperature $T(r)$. The radial profiles of hot gas is then
\begin{equation}\label{eq:rho2ne}
\rho_{\rm hot}(r) = \bar \mu {m_{\rm{H}}}n_{e}(r),
\end{equation}
where $\bar \mu {m_{\rm{H}}}$ is the mean particle mass.

We save the results $\rho_{\rm hot}(r_i)$, $n_e(r_i)$, $T(r_i)$ in each shell of radius $r_i$ given by Eq. (\ref{eq:radial}) at all times. In this paper, we do not consider the radial distribution of metallicity in the hot gaseous phase for gas cooling and chemical enrichment by supernova. We simply assume homogeneous value for the hot gas metallicity $Z_{\rm hot}$ in each halo.


\subsection{Gas cooling and infall} \label{sec:coolingmodel}

According to White \& Rees (1978), White \& Frenk (1991) and Springel et al. (2001), the hot gas is spherically distributed in each dark matter halo. The local cooling time at a given radius $r$ is the ratio between the thermal energy density of the hot gas and the cooling rate in a unit volume
\begin{equation}\label{eq:tcool}
{t_{{\rm{cool}}}}\left( r \right) = \frac{{3\bar \mu {m_{\rm{H}}}{k_{\rm{B}}}T}}{{2{\rho _{{\rm{hot}}}}\left( r \right)\Lambda \left( {T,Z} \right)}},
\end{equation}
where $\bar \mu {m_{\rm{H}}}$ is the mean particle mass, $k_{\rm B}$ is the Boltzmann constant, $\rho_{\rm{hot}}(r)$ is the radial density profile of the hot gas, $T$ and $Z$ are the temperature and metallicity of hot gas, and $\Lambda \left( {T,Z} \right)$ is the cooling function for collisional cooling processes. According to Sutherland \& Dopita (1993), when the hot gas is in collisional ionization equilibrium, the cooling function depends only on the metallicity and temperature, but not on its density.

The cooling recipes compare $t_{\rm cool}$ with the dynamic time scale of the halo $t_{\rm dyn}=r_{\rm 200}/v_{\rm 200}=0.1H(z)^{-1}$ (De Lucia et al. 2004), and we define the cooling radius where the cooling time equals to the dynamic time scale of the halo. If $r_{\rm cool}>r_{\rm 200}$, the cooling is in ``fast mode'', and the hot gas inside the whole halo will be accreted to the central galaxy in one dynamical time scale $t_{\rm dyn}$. This process is called a ``cold flow'', and the cooling rate for rapid cooling is
\begin{equation}\label{eq:rapidcool}
\dot m_{\rm{cool}} = {m_{{\rm{hot}}}}/{t_{{\rm{dyn}}}}.
\end{equation}
For haloes with $r_{\rm cool}<r_{\rm 200}$, cooling is in the ``slow mode'', and only the hot gas inside the cooling radius is accreted onto the central galaxy in one $t_{\rm dyn}$. The cooling rate of this quasi-static cooling can be written as
\begin{equation}\label{eq:coolgeneral}
{\dot m_{{\rm{cool}}}} = \frac{{{m_{{\rm{hot}}}}\left(<{{r_{{\rm{cool}}}}} \right)}}{{{t_{{\rm{dyn}}}}}},
\end{equation}
where $m_{\rm hot}\left(<r_{{\rm{cool}}} \right)$ represents the mass of hot gas within radius $r_{\rm cool}$. For hot gas in an isothermal distribution, Eq. \ref{eq:coolgeneral} can be written as
\begin{equation}\label{eq:coolisothermal}
{\dot m_{{\rm{cool}}}} = {m_{{\rm{hot}}}}\frac{{{r_{{\rm{cool}}}}}}{{{r_{{\rm{200}}}}}}\frac{1}{{{t_{{\rm{dyn}}}}}}
\end{equation}

\begin{figure}
\centering
 \includegraphics[angle=0,scale=0.45]{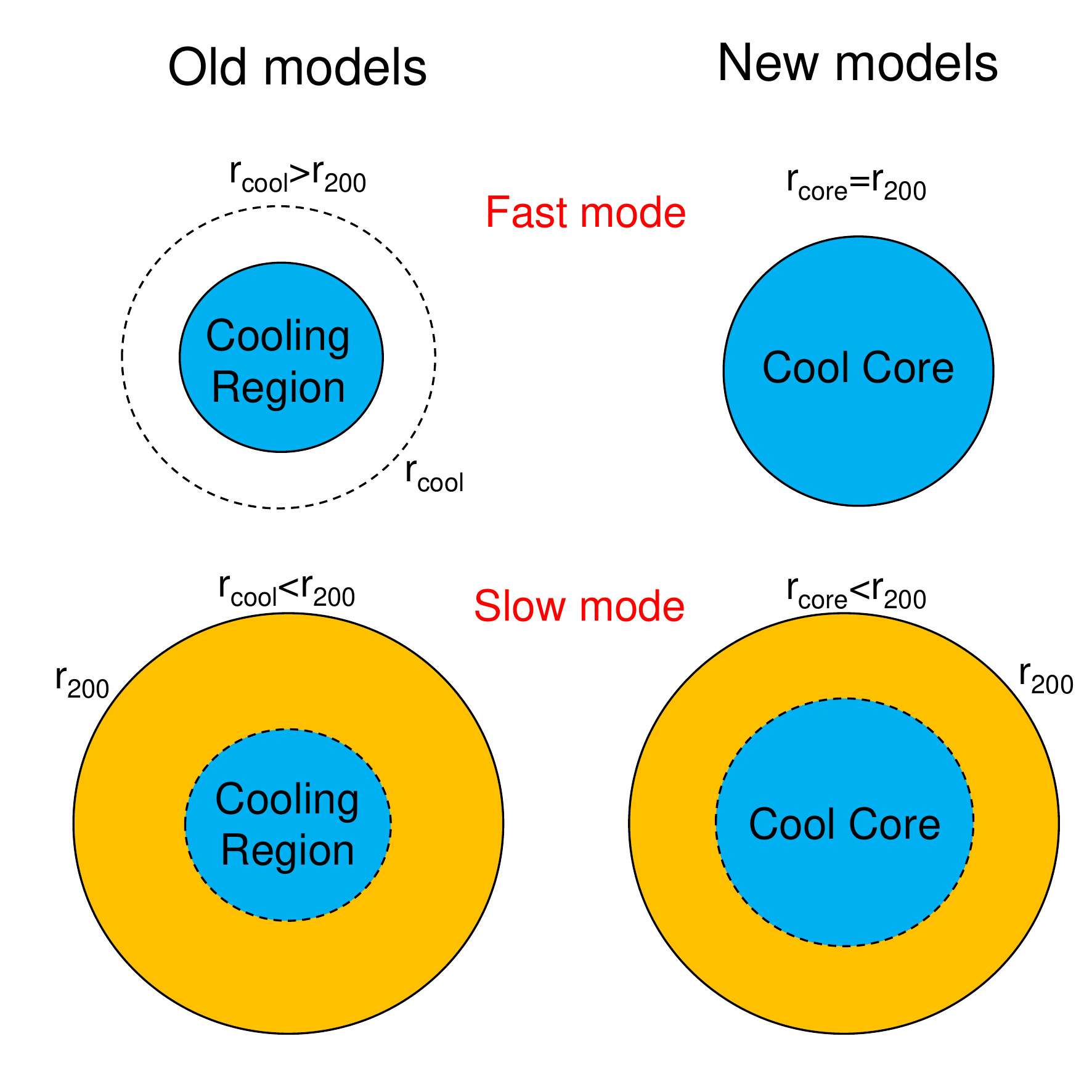}
 \caption{The comparison of the cooling recipes between the old and new models. The blue regions represent the cooling regions or cool cores, and the yellow regions represent the hot gaseous halo in stable state without condensation and infall.}\label{fig:coolmodel}
\end{figure}

In the new model, the gas cooling and infall recipes should be changed a bit according to the new profile of the hot gas. In Fig. \ref{fig:coolmodel}, we show an illustration of the cooling recipes in the old and new models. The main difference is the radius of the cooling region, and we adopt the cool core instead of a cooling region in the new model. The cool core radius $r_{\rm core}$ in the new model is analogous to the cooling radius in the previous model, and the gas inside the cool core radius is assumed to form cold filaments because of the local thermal instability. These filaments fall to the central galaxy in the halo. The infall time-scale of a cool core is defined as the free-fall time scale at the core radius,
\begin{equation}\label{eq:tinfall}
t_{\rm infall}=r_{\rm core}/v_{\rm 200}.
\end{equation}
The ``fast mode'' in the new model corresponds to haloes with $t_{\rm TI}/t_{\rm ff}=10$ at the virial radius, and the whole halo should be regarded as a cool core. 
The ``slow mode'' in the new model corresponds to the ``cool-core cluster'', \ie{}the halo with $r_{\rm core}<r_{\rm 200}$, and the hot gas inside cool core will infall in one $t_{\rm infall}$. Similar to Eq. \ref{eq:coolgeneral}, the infall rate can be written as
\begin{equation}\label{eq:coolcorerate}
{\dot m_{{\rm{cool}}}} = \frac{{{m_{{\rm{hot}}}}\left(<{{r_{{\rm{core}}}}} \right)}}{{{t_{{\rm{infall}}}}}},
\end{equation}
where $m_{\rm hot}\left(<r_{\rm{core}}\right)$ is the total hot gas mass in a cool core.

\begin{figure}
\centering
 \includegraphics[angle=0,scale=0.4]{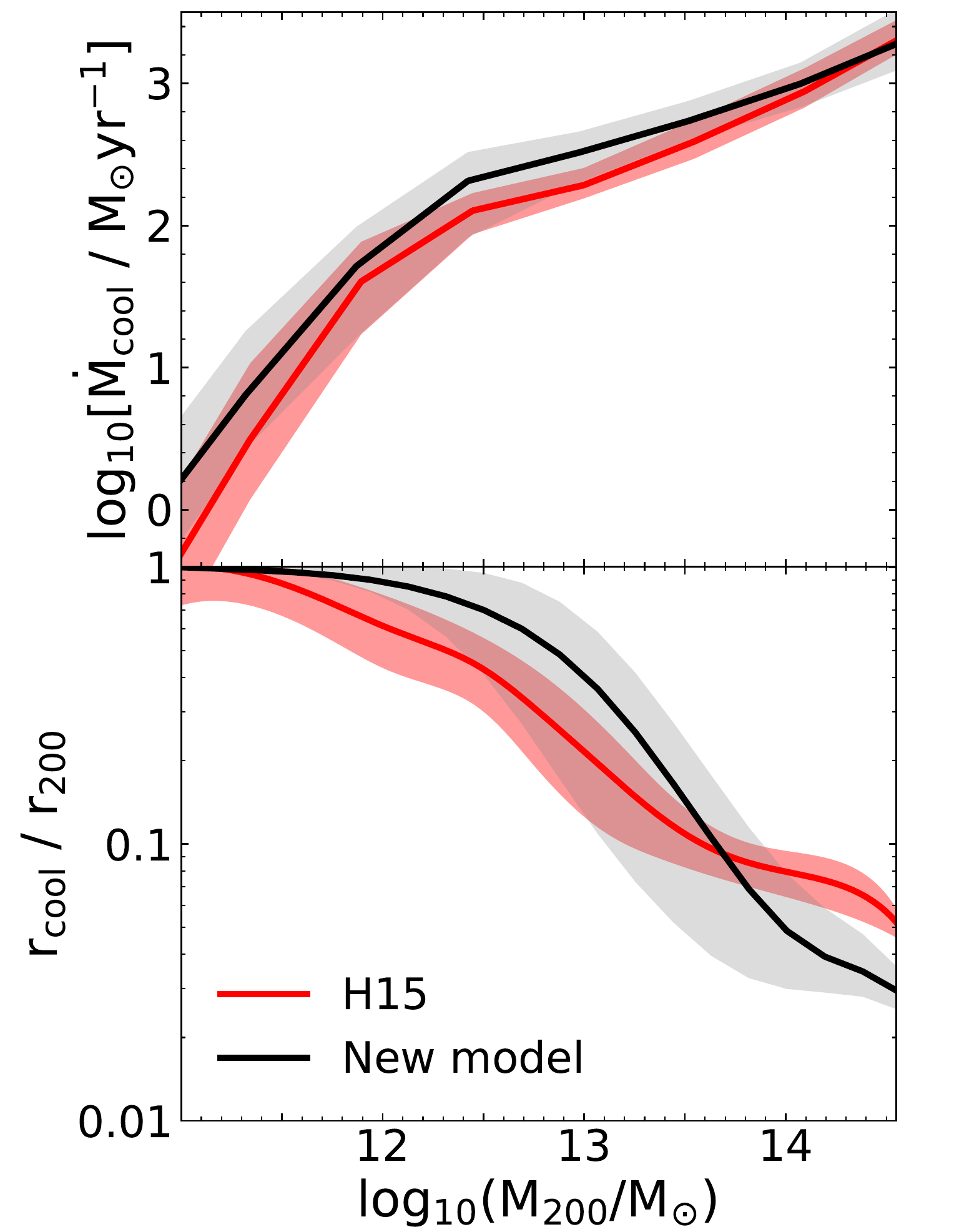}
 \caption{The comparison of the cooling rate (top panel, without the suppression by radio-mode AGN) and the cooling radius (bottom panel) in haloes with different mass from the old (H15) and our new models at $z=0$. The solid curves are the mean values for the haloes with central galaxies (Type 0 \& 1 in the models), and the shaded areas represent $\pm 1\sigma$ deviations for the model samples.}\label{fig:rcoolcompare}
\end{figure}
In Fig. \ref{fig:rcoolcompare}, we show the comparison of the cooling rate and cooling radius in haloes with different masses from the old and new models. The results from both models are very similar, and the proportion of the small haloes in fast mode cooling is only a bit higher in the new model. The main reason for the similar cooling rates from the two models is that we have readjusted the model parameters in the new model so that both the new and old models fit the same observational results like the redshift evolution of cosmic star formation density and the mass functions at $z=0$ (see Fig. \ref{fig:massfunction} \& \ref{fig:sfrredshift} in Sec. \ref{sec:sfmodel}). Therefore, we expect that the galaxy properties from the new cooling model should not differ much from the old model.

\subsection{Reheating and ejection by supernova feedback} \label{sec:snmodel}

In H15 model, the instantaneous recycling approximation is assumed, \ie{}low mass stars keep the stellar mass eternally and high mass stars explode as supernovae at the time step of their formation. If the energy released by supernovae exceeds the thermal energy of the reheated hot gas, part of the hot gas can be ejected from the halo. The energy from supernovae can be written as
\begin{equation}\label{eq:snenergy}
\Delta E_{\rm SN}=0.5\epsilon_{\rm halo}\Delta m_* v_{\rm SN}^2
\end{equation}
where $\Delta m_*$ is the mass of newly formed star and $0.5v_{\rm SN}^2$ is the energy of supernova ejecta per unit mass of newly formed stars, and $v_{\rm SN}$ is a model parameter representing the velocity of SN ejecta. The supernova energy can reheat part of the cold gas in the ISM disk to the hot gas in halo, and $\epsilon_{\rm halo}$ in Eq. \ref{eq:snenergy} is a halo-dependent parameter representing the efficiency of the supernovae converting the ISM cold gas to the halo hot gas. The mass of cold gas reheated by supernovae in a given time step is \begin{equation}\label{eq:reheat}
\Delta m_{\rm reheat}=\epsilon_{\rm disk}\Delta m_*,
\end{equation}
where $\epsilon_{\rm disk}$ is a disk-dependent parameter representing the supernova reheating efficiency. According to Guo et al. (2011) and H15, both $\epsilon_{\rm disk}$ and $\epsilon_{\rm halo}$ are related to the properties of dark matter halo
\begin{equation}\label{eq:epsilonguo}
\begin{array}{l}
\epsilon_{\rm disk}=\epsilon\left[ {0.5 + {{\left( {\frac{{{v_{{\rm{max}}}}}}{{{v_{{\rm{reheat}}}}}}} \right)}^{ - {\beta _1}}}} \right]\\
\epsilon_{\rm halo}=\eta\left[ {0.5 + {{\left( {\frac{{{v_{{\rm{max}}}}}}{{{v_{{\rm{eject}}}}}}} \right)}^{ - {\beta _2}}}} \right]
\end{array}
\end{equation}
and $\epsilon$, $\eta$, $v_{\rm reheat}$, $v_{\rm eject}$, and $\beta_1=\beta_2$ are model parameters related to the galaxy disk and halo. In previous models, the reheated cold gas becomes the halo hot gas in isothermal profiles with $T_{\rm hot}=T_{\rm 200}$, and the energy of the cold gas reheated by supernovae is
\begin{equation}\label{eq:reheatenergy}
\Delta {E_{{\rm{reheat}}}} = \frac{1}{2}\Delta {m_{{\rm{reheat}}}}v_{{\rm{200}}}^2.
\end{equation}
If the energy released by supernovae (Eq. \ref{eq:reheat}) exceeds the energy stored in the halo hot gas (Eq. \ref{eq:reheatenergy}), gas can be ejected from the halo (c.f. Eq. \ref{eq:meject}).

In the new model, the halo hot gas is no longer isothermal, and the temperature of the hot gaseous halo should change because of the mass and energy injection by supernova reheating. We assume $T_{\rm hot}^{-}$ and $T_{\rm hot}^{+}$ as the hot gas temperature before and after supernova reheating, and we can get the radial distribution of $T_{\rm hot}^{-}$ and $T_{\rm hot}^{+}$ from the Sharma12 model combined with the mass of reheated gas in Eq. \ref{eq:reheat}. Then, the energy injected into the hot gas halo during the supernova reheating can be written as
\begin{multline}\label{eq:ereheat}
\Delta {E_{{\rm{reheat}}}} = \\
\frac{{3{k_{\rm{B}}}}}{{2\bar \mu {m_{\rm{H}}}}}\left[ {\Delta {m_{{\rm{reheat}}}}\bar T_{{\rm{hot}}}^ +  + \left( {\bar T_{{\rm{hot}}}^ +  - \bar T_{{\rm{hot}}}^ - } \right)m_{{\rm{hot}}}^ - } \right],
\end{multline}
where ${\Delta {m_{{\rm{reheat}}}}/\bar \mu {m_{\rm{H}}}}$ and $\bar T_{\rm hot}$ are the particle numbers and mean temperature of the hot gas in the whole halo. The superscript ``-'' and ``+'' refer to the quantities before and after mass/energy injection. The first term ${\Delta {m_{{\rm{reheat}}}}\bar T_{{\rm{hot}}}^ + }$ represents the thermal energy of the reheated cold gas, and the second term ${\left( {\bar T_{{\rm{hot}}}^ +  - \bar T_{{\rm{hot}}}^ - } \right)m_{{\rm{hot}}}^-}$ represents the energy change of the whole gas halo. We assume the temperature of the reheated gas equals to the mean temperature of the halo after reheating $T_{\rm hot}^+$, or the reheated gas has the same specific thermal energy as the halo hot gas.

In Eq. \ref{eq:ereheat}, the mass-weighted mean temperature of the hot gas $T_{\rm hot}^-$ and $T_{\rm hot}^+$ can be calculated by
\begin{equation}\label{eq:treheat}
\bar{T}_{\rm hot} = m_{{\rm{hot}}}^{-1}\int_0^{{r_{\rm 200}}} {T\left( r \right){\rho _{{\rm{hot}}}}\left( r \right)} 4\pi {r^2}dr
\end{equation}
Combining Eq. \ref{eq:ereheat} \& \ref{eq:treheat}, we can get $\Delta E_{\rm{reheat}}$.

Following the prescription in H15 model, if $\Delta E_{\rm SN}>\Delta E_{\rm{reheat}}$, the excess energy from supernovae $\Delta E_{\rm SN}-\Delta E_{\rm{reheat}}$ can eject part of the hot gas out of the halo potential well, and the total mass of ejected hot gas is
\begin{equation}\label{eq:meject}
\Delta m_{\rm eject}=\frac{{\Delta {E_{{\rm{SN}}}} - \Delta {E_{{\rm{reheat}}}}}}{{\frac{1}{2}v_{{\rm{200}}}^2}}.
\end{equation}
In the current model, we do not consider which part of the hot gas is ejected, and simply assume same proportion is ejected at all the radius of the hot gas halo.

\begin{figure}
\centering
 \includegraphics[angle=0,scale=0.4]{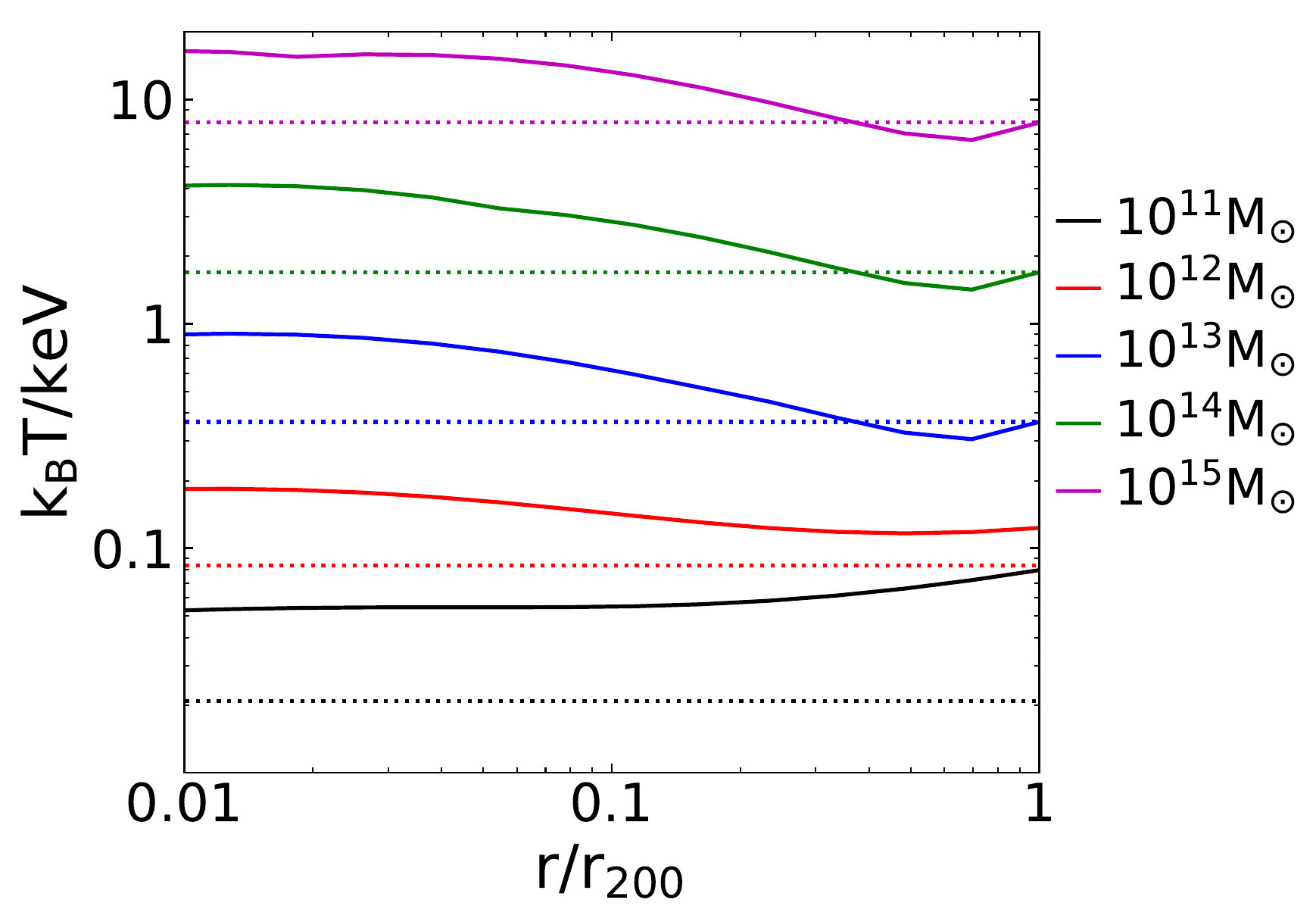}
 \caption{Radial profiles of the hot gas temperature from Sharma12 models (solid curves) and the isothermal spheres (dotted curves). Similar to Fig.\ref{fig:neprofiles}, the curves in different colors represent the results from different halo masses, and the model input parameters are the same as  in Fig.\ref{fig:neprofiles}.
 }\label{fig:tprofiles}
\end{figure}

In Fig. \ref{fig:tprofiles}, we compare the radial profiles of gas temperature $T_{\rm hot}(r)$ from our new model with the virial temperature $T_{\rm 200}$ adopted in isothermal model. We can see that $T_{\rm hot}(r)$ in the new models is higher than $T_{\rm 200}$ in most regions of haloes with different mass. Thus, the reheated temperature $T_{\rm reheat}$ and energy $\Delta E_{\rm reheat}$ in Eq. \ref{eq:ereheat} \& \ref{eq:treheat} in the new model tend to be higher than isothermal models, and less hot gas in the halo is ejected by the excess energy from supernovae.

The hot gas ejected by supernovae becomes external baryonic reservoir around dark matter haloes, and will be reincorporated to the halo again at later time with the growth of the halo. Following H15 paper, the reincorporation rate is inversely proportional to the mass of the host halo,
\begin{equation}\label{eq:reincorporate}
{\dot m_{{\rm{rein}}}} = - {\left[ {{\gamma _{{\rm{rein}}}}\frac{{{{10}^{10}}{M_ \odot }}}{{{m_{{\rm{200}}}}}}} \right]^{ - 1}}{m_{{\rm{ejected}}}},
\end{equation}
where $\left(\gamma_{\rm rein}10^{10}\ms/m_{\rm 200}\right)$ is the reincorporation time scale and $\gamma _{{\rm{rein}}}$ is an adjusted efficiency; note that the reincoporation time is shorter for a more massive halo. In the reincorporation process, we do not trace the exact location of the hot gas reincorporated into the halo, and we simply add the mass of reincorporated gas to total hot gas mass and recalculate the radial profiles of the hot gas before we process the cooling and infall recipes.

\subsection{AGN feedback} \label{sec:AGN}

A black hole can be embedded in the centre of a galaxy, which grows by accreting cold gas from the ISM or hot gas from the halo. In the H15 model, cold gas can be accreted by the black hole during mergers via the ``quasar mode'' and hot gas 
quiescently from the hot gaseous halo via the ``radio mode''. Radio-mode accretion pumps energy into the hot gas and suppresses gas cooling in massive galaxies with large black holes. 
This AGN feedback further reduces the net cooling rate, preventing rapid cooling in the centres of massive haloes.

In the model, the radio-mode accretion rate from hot gas (taken to be proportional to the black hole and halo masses) is
\begin{equation}\label{eq:radiomode}
\dot m_{\rm{BH}}=\kappa_{\rm AGN}\left({\frac{{m_{{\rm{BH}}}}}{{{{10}^8}\ms}}} \right)\left( {\frac{{{f_{{\rm{hot}}}}}}{{0.1}}} \right){\left( {\frac{{{v_{{\rm{200}}}}}}{{200~{\rm{km~s}^{-1}}}}}\right)^3},
\end{equation}
where $m_{\rm BH}$ is the black hole mass, $f_{\rm hot}$ is the 
hot gas mass fraction, and $\kappa_{\rm AGN}$ is the characteristic accretion efficiency. 
The energy pumped back into the hot gas by radio mode feedback 
is
\begin{equation}\label{eq:eddington}
L_{\rm BH}=0.1\dot m_{\rm{BH}}c^2,
\end{equation}
where $c$ is the speed of light and the 
feedback efficiency is set to 0.1. The energy produced by accretion offsets cooling, returning a net mass cooling rate of
\begin{equation}\label{eq:netcoolingrateold}
\dot m_{{\rm{cool}}}'=\dot m_{\rm cool}-\frac{{{L_{{\rm{BH}}}}}}{{\frac{1}{2}v_{{\rm{200}}}^2}},
\end{equation}
where we only keep the non-negative value of $\dot m_{{\rm{cool}}}'$. 

In the H15 model, $\frac{1}{2}v_{\rm{200}}^2$ in Eq. \ref{eq:netcoolingrateold} is the thermal energy per unit mass of isothermal hot gas 
with $T_{\rm hot}=T_{\rm 200}$. In the new model, we adopt $\bar{T}_{\rm hot}$ in Eq. 
\ref{eq:treheat} as the mean temperature of the halo hot gas, and thus the net cooling rate can be written as
\begin{equation}\label{eq:netcoolingratenew}
\dot m_{{\rm{cool}}}' = {\dot m_{{\rm{cool}}}} - \frac{{{L_{{\rm{BH}}}}}}{{3{k_{\rm{B}}}{{\bar T}_{{\rm{hot}}}}/2\bar \mu {m_{\rm{H}}}}},
\end{equation}
where $\bar{\mu}m_{\rm H}$ is the mean particle mass and $k_{\rm B}$ is the Boltzmann constant. In practice, this radio-mode feedback model completely shuts down cooling flows onto massive early-type galaxies, even those which are the brightest central galaxy (BCG) of a ``cool core cluster'' as defined in Sec. \ref{sec:coolingmodel}.

\subsection{Gas stripping by tidal force and ram pressure} \label{sec:stripping model}

Satellite galaxies move within the gravitational potential of their central halo. This motion within the gaseous ICM of galaxy clusters can affect the evolution of satellites. In L-Galaxies, this environmental effect includes gas stripping by ram pressure from the ICM and the tidal force of the central galaxy. In this section, we will describe how we change the prescriptions of ram pressure and tidal stripping based on the new hot gas profiles.

Following the paper by Guo et al. (2011), when a smaller halo falls into 
a larger halo, the tidal force will gradually remove its dark matter. The smaller halo becomes a satellite and it stops accreting baryonic matter from the environment. The hot gas around the satellite galaxy is stripped together with the dark matter halo. In previous models, hot gas is assumed to follow the density profile of the dark matter halo, and the hot gas fraction of the satellite experiencing tidal stripping remained constant,
\begin{equation}\label{eq:tidalold}
\frac{{{m_{{\rm{hot, infall}}}}}}{{{m_{{\rm{DM, infall}}}}}} = \frac{{{m_{{\rm{hot}}}}\left( {{r_{{\rm{tidal}}}}} \right)}}{{{m_{{\rm{DM, tidal}}}}}},
\end{equation}
where ${{m_{{\rm{hot, infall}}}}}$ and ${{m_{{\rm{DM, infall}}}}}$ are the hot gas and dark matter masses of the halo just before the infall. Here, ${{m_{{\rm{hot}}}}\left( {{r_{{\rm{tidal}}}}} \right)}$ and ${{m_{{\rm{DM, tidal}}}}}$ are the hot gas and dark matter masses of the current satellite halo, and the hot gas at a radius larger than $r_{\rm tidal}$ is stripped by the tidal force. For the isothermal sphere, ${r_{{\rm{tidal}}}} = {r_{{\rm{DM, infall}}}}\left( {{m_{{\rm{DM}}}}/{m_{{\rm{DM, infall}}}}} \right)$. 

In the new model, the dark matter halo and hot gas follow different profiles, and we define a new tidal stripping radius $r_{\rm tidal}$ for the satellite galaxy. We assume that the current dark matter mass of a satellite equals the dark matter mass inside $r_{\rm tidal}$ just before the infall,
\begin{equation}\label{eq:tidal}
m_{\rm{DM, tidal}}=m_{\rm DM, infall} (r<r_{\rm tidal}).
\end{equation}
Based on the NFW profile (Eq. \ref{eq:genNFW}), we sum the dark matter halo mass in each shell from inside to outside. The radius where the dark matter mass inside the shell equals $m_{\rm{DM, tidal}}$ in Eq. \ref{eq:tidal} is taken as the tidal radius $r_{\rm tidal}$.



The other environmental effect is ram pressure. When a satellite galaxy is moving through the gravitational potential of a galaxy cluster, the hot gas in the ICM exerts a pressure on the satellite. If the gravity of the satellite galaxy is smaller than the ram pressure exerted by the host halo, the hot gas of the satellite begins to be stripped. We assume $r_{\rm rp}$ as the radius from the centre of the satellite where the self-gravity is approximately balanced by the ram pressure:
\begin{equation}\label{eq:rampressure}
{\rho_{{\rm{hot, sat}}}}({r_{{\rm{rp}}}})v_{{\rm{sat}}}^2 = {\rho _{{\rm{hot, cen}}}}(r)v_{{\rm{orbit}}}^2,
\end{equation}
where $\rho_{{\rm{hot, sat}}}({r_{{\rm{rp}}}})$ is the hot gas density of satellite at radius $r_{\rm rp}$, $v_{{\rm{sat}}}$ is the virial velocity of the subhalo, ${\rho _{{\rm{hot, cen}}}}(r)$ is the hot gas density profile of the host halo and $r$ is the distance from the host halo centre to the satellite, and $v_{{\rm{orbit}}}$ is the orbital velocity of the satellite relative to the host centre (which we assume to be equal to $v_{\rm 200}$ of the host halo). Here, $\rho_{{\rm{hot, sat}}}(r)$ and ${\rho _{{\rm{hot, cen}}}}(r)$  are the radial profiles of the hot gas in the haloes of satellite and host, and the new radial gas density model  affects the calculation of $r_{\rm rp}$ in Eq. \ref{eq:rampressure}.

In real galaxies, ram pressure can also strip the cold gas in the ISM. For example, recent observational results from Wang et al. (2021) show the ram pressure stripping of  HI gas in a galaxy cluster. The SAM work by Luo et al. (2016) models the ram pressure stripping of the ISM and the effect on star formation due to the stripping. They find that the ram pressure stripping of the cold gas mainly affects the star formation in low-mass satellites located in massive galaxy clusters. None the less, in this paper we only consider the ram pressure effect on the hot gas. 

Once $r_{\rm tidal}$ and $r_{\rm{rp}}$ are calculated from Eq. \ref{eq:tidal} \& \ref{eq:rampressure}, the minimum of the two values is set as the stripping radius $r_{\rm strip}$,
\begin{equation}\label{eq:rstrip}
r_{\rm strip}=\min(r_{\rm{tidal}}, r_{\rm{rp}}),
\end{equation}
and the hot gas of the subhalo beyond $r_{\rm strip}$ is stripped and added to the host halo.

\subsection{Modelling cold gas and star formation in ISM} \label{sec:sfmodel}

In this paper, we also consider the results of atomic and molecular gas components of the ISM to constrain the models with observations in the nearby Universe. In some versions of L-Galaxies (\eg{}Fu et al. 2013; Luo et al. 2016; Henriques et al. 2020; Yates et al. 2021a), various physical prescriptions trace the radial distribution of stars and cold gas in each galaxy disc. In this paper, the main purpose is to study the hot gaseous halo, so we follow the model version from H15, which does not include radially-resolved discs. We adopt a simplified scheme to include the molecular and atomic phases in the ISM, \ie{}we only calculate the total mass of HI and $\h2$ gas in each galaxy.

In each galaxy disk, we simply assume both stellar and cold gas components to be distributed in an exponential profile
\begin{equation} \label{eq:exponential}
{\Sigma _{*,{\rm{gas}}}}(r) = \Sigma _{*,{\rm{gas}}}^0\exp \left( - r/{r_{*,{\rm{gas}}}} \right),
\end{equation}
where the subscripts ``$*$'' and ``gas'' represent star and cold gas components respectively. In each exponential disk, $\Sigma_{*,{\rm gas}}^0 = {m_{*,{\rm{gas}}}}/2\pi r_{*,{\rm{gas}}}^2$ represents the central surface density and $r_{*,{\rm{gas}}}$ represents the scale length of the gas or stellar disk, which is calculated from the ratio of the specific angular momentum of the disk and the maximum circular velocity of the dark matter halo, \ie{}${r_{*,{\rm{gas}}}} = {j_{*,{\rm{gas}}}}/2{v_{{\rm{max}}}}$. 

We calculate the partition between atomic and molecular gas in the ISM with a prescription similar to the models of BR06 (Blitz \& Rosolowsky 2006) adopted in Fu et al. (2010, 2013), with the molecular-to-atomic fraction $R_{\rm H_2}$ related to the interstellar pressure as follows
\begin{equation}\label{eq:br06}
{R_{{{\rm{H}}_{\rm{2}}}}}\left( r \right) = \frac{{{m_{{{\rm{H}}_2}}}\left( r \right)}}{{{m_{{\rm{HI}}}}\left( r \right)}} = {\left[ {\frac{{P\left( r \right)}}{{{P_0}}}} \right]^{{\alpha _P}}},
\end{equation}
where $P_0$ and $\alpha_P$ are free parameters (see Table 1). In the midplane of a galaxy disk, the ISM pressure $P(r)$ in Eq. \ref{eq:br06} is given by
\begin{equation}\label{eq:pressure}
P(r)=\frac{\pi}{2}G\Sigma_{\rm gas}\left(r\right)\left[\Sigma_{\rm gas}\left(r\right)+0.1\sqrt{\Sigma_*\left(r\right)\Sigma_*^0}\right],
\end{equation}
where $\Sigma_{\rm gas}(r)$, $\Sigma_*(r)$ and $\Sigma_{*}^0$ are from the exponential radial profiles in Eq. \ref{eq:exponential}.

In Eq. \ref{eq:br06} \& \ref{eq:pressure}, we assume the $\h2$ and HI gas to include the mass of elements heavier than hydrogen, i. e.,  $m_{\rm cold}=m_{\rm HI}+m_{\h2}$. 


In the H15 model, the star formation law follows the models in De Lucia \& Blaizot (2007) and Guo et al. (2011)
\begin{equation}\label{eq:oldsfr}
{\dot m_*} = \frac{{{\alpha _{{\rm{SF}}}}}}{{{t_{{\rm{dyn}}}}}}\left( {{m_{{\rm{gas}}}} - {m_{{\rm{crit}}}}} \right)
\end{equation}
and the star formation rate is related to total mass of the cold gas $m_{\rm gas}$, critical mass of a gas disk $m_{\rm crit}$, and the dynamic time scale of the disk $t_{\rm dyn}$. According to recent observational results, \eg{}from SCUBA (Bourne et al. 2017; Schruba et al. 2011), the star formation rate should be related to the mass of molecular gas (CO or $\h2$) rather than the total cold gas. Therefore, following the model work in Fu et al. (2013) and some recent works like Xie et al. (2017), we adopt a star formation law proportional to the mass of $\h2$ gas in the ISM for our new model, \ie{}
\begin{equation}\label{eq:h2sfr}
{\dot m_*} =\alpha_{\h2} m_{\h2},
\end{equation}
where $m_{\h2}$ is the mass of gas in the molecular phase, and the parameter $\alpha_{\h2}$ is the molecular gas star formation efficiency which represents the depletion time scale of the molecular gas by star formation.

\subsection{The parameters and calibrations of the model}

\begin{table*}
    \centering
    \caption{The model parameters. The top section shows the parameters introduced or changed in this paper, and the bottom section shows the parameters remaining the same values as in previous papers.}\label{tab:parameters}
    \begin{tabular}{l|l|l}
     \hline \hline
      Parameter & Value& Description\\
      \hline
      $ \alpha_{\h2}$          &  $5.7\times10^{-10}\rm{yr}^{-1}$ & molecular gas star formation efficiency \\
      $\alpha_{\rm P}$         &  0.8   & index in the relation between molecular fraction and ISM pressure \\
      $P_0$                    &  $5.93 \times 10^{13}$ Pa & constant in the relation between molecular fraction and ISM pressure \\
      $\kappa_{\rm AGN}$       &  $1.2\times 10^{-2}\ms\rm{yr}^{-1}$ & quiescent black hole accretion rate\\
      $f_{\rm BH}$             &  0.135 & cold gas BH accretion fraction in quasar mode\\
      $\epsilon_{\rm reheat}$  &  1.5   & amplitude of SN reheating efficiency\\
      $\beta_{\rm reheat}$     &  1.2   & SN reheating efficiency slope\\
      $\eta_{\rm eject}$       &  0.65  & amplitude of SN ejection efficiency\\
      $\beta_{\rm eject}$      &  0.8   & SN ejection efficiency slope\\
      $\gamma_{\rm reinc}$     &  $0.85\times10^{10}\rm{yr}^{-1}$ & reincorporation efficiency\\
      $R_{\rm merger}$         &  0.2   &  satellite ratio threshold between major and minor merger \\
      $s$                      &  3.0   &  asymptotic gas density slope in Eq. \ref{eq:genNFW}\\
      \hline
      $\alpha_{\rm SF, burst}$ &  1.9  & index in merger induced starburst \\
      $\beta_{\rm SF, burst}$  &  0.6  & constant in merger induced starburst \\
      $v_{\rm BH}$             &  750 km~s$^{-1}$ & normalization of BH accretion in quasar mode\\
      $v_{\rm reheat}$         &  480 km~s$^{-1}$ & normalization of SN reheating efficiency\\
      $v_{\rm eject}$          &  100 km~s$^{-1}$ & normalization of SN ejection efficiency\\
      $m_{\rm r.p.}$           &  $1.2\times10^{14}\ms h^{-1}$ & threshold halo mass for ram pressure stripping\\
      $Y$                      &  0.046 & fraction of metals instantaneously returned after star formation \\
      $R$                      &  0.43  & fraction of star formation mass instantaneously recycled back to cold gas \\
    \hline \hline
    \end{tabular}
\end{table*}

As we have changed the physical prescriptions mentioned above, it is necessary to recalibrate the model parameters to fit the data from various observations. In Tab. \ref{tab:parameters}, we list the parameters used in our models. The top section of Tab. \ref{tab:parameters} includes the parameters we tune to fit the observations, and the bottom section includes the parameters retaining  the same values as in previous models.

\begin{figure*}
\centering
 \includegraphics[angle=0,scale=0.4]{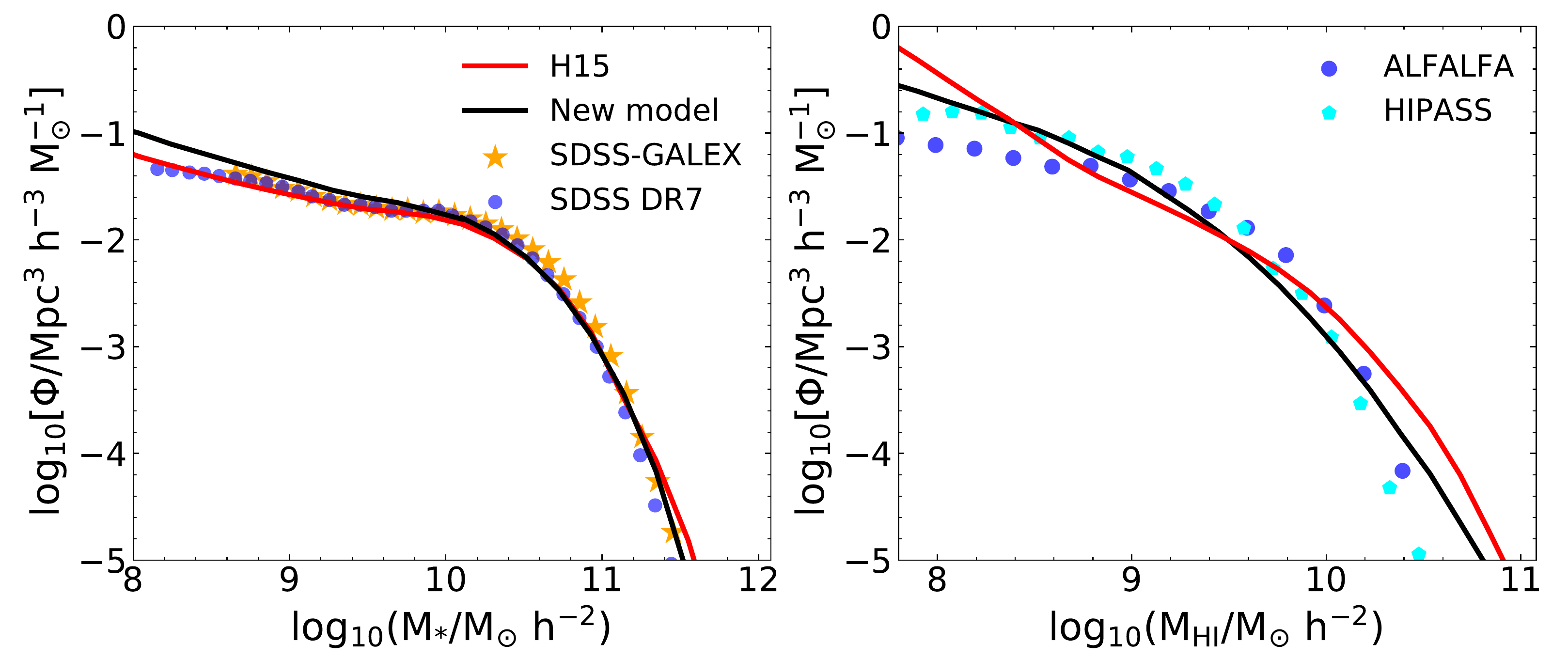}
 \caption{Left panel: the stellar mass function at $z=0$. The red and black curve are from H15 model and our new model respectively. The data points are from SDSS DR7 (Li \& White 2009) and SDSS-GALEX (Moustakas et al. 2013). Right panel: the HI mass function at $z=0$. We show the model results and compare with the observations from HIPASS (Zwaan et al. 2005) and ALFALFA (Jones et al. 2018). The black curve is from the new model and the red curve is from H15 model by assuming $m_{\rm HI}=0.54m_{\rm cold~gas}$.
 }\label{fig:massfunction}
\end{figure*}

\begin{figure}
\centering
 \includegraphics[angle=0,scale=0.4]{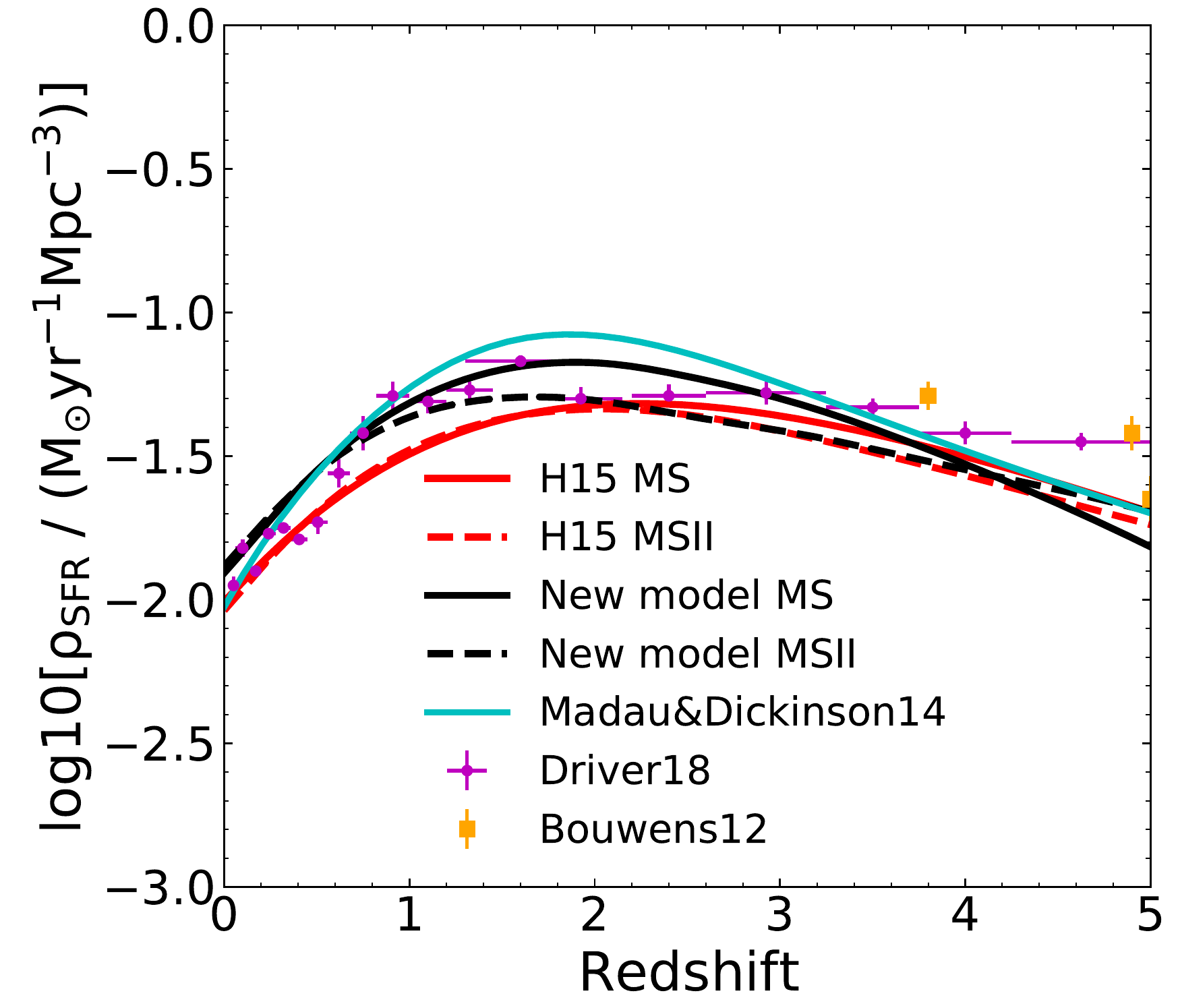}
 \caption{The redshift evolution of cosmic star formation density: the red and black curves represent the results based on the new model and previous H15 model respectively (solid lines based on MS and dashed lines based on MS-II). The cyan curve is from the fitting results in Madau \& Dickinson (2014) by assuming a Chabrier IMF. The data points are from Bouwens et al. (2012a,b) and Driver et al. (2018).} 
 \label{fig:sfrredshift}
\end{figure}

We tune the parameters to fit: (1) The stellar mass function and HI mass function at $z=0$; (2) The redshift evolution of cosmic star formation density; (3) The scaling relations of X-ray luminosity at $z=0$ (e. g., $\LX-M_{\rm halo}$, $\LX-\TX$ relations). In Fig. \ref{fig:massfunction}, we show the model results for the stellar mass function and HI mass function at $z=0$, and compare with the old models and observations. We combine both the MS and MS-II to extend the mass functions to the low mass range. Since the H15 model does not contain the partition between HI and $\h2$ gas, we 
assume $m_{\rm HI}=0.54m_{\rm cold~gas}$ for simplicity.

Fig. \ref{fig:massfunction} shows the close similarity between the stellar mass functions from the H15 and new models, at least above $\textnormal{log}(M_{*}/\ms) \sim{} 10.0$, and the new model's improved agreement with observations of the HI mass function.
Fig. \ref{fig:sfrredshift} shows the redshift evolution of the cosmic star formation rate density, compared to observations from Bouwens et al. (2012a,b), Madau \& Dickinson (2014), and Driver et al. (2018). We can see that the new model is a better match to observations than the H15 model. 

The main reason for the difference is not the change of cooling prescription but the change of star formation prescription in Eq. \ref{eq:h2sfr}. The molecular star formation efficiency $\alpha_{\h2}$ in Eq. \ref{eq:h2sfr} is a constant independent of redshift, while the previous star formation prescription Eq. \ref{eq:oldsfr} is proportional to $t_{\rm dyn}^{-1}$. According to the discussion in Fu et al. (2012), the dynamic time scale $t_{\rm dyn}^{-1}$ in the star formation law can increase the star formation rate at high redshifts because $t_{\rm dyn}^{-1}\sim{\left( {1 + z} \right)^{3/2}}$.

\begin{figure*}
\centering
 \includegraphics[angle=0,scale=0.8]{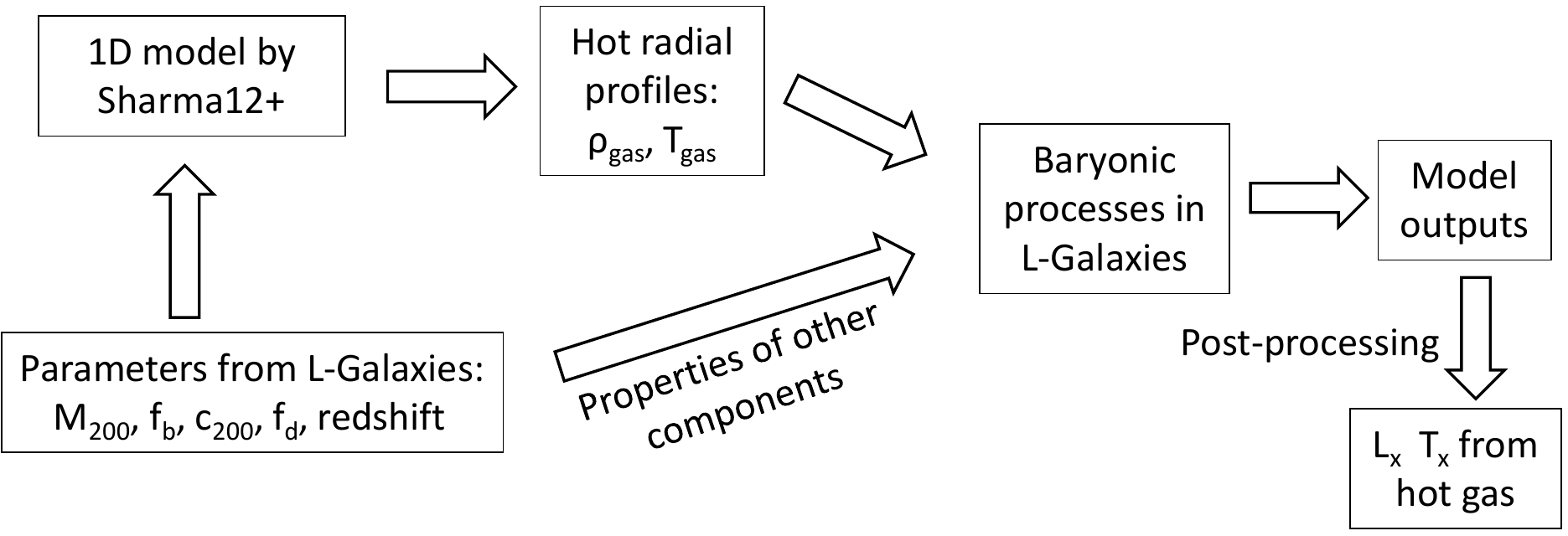}
 \caption{The brief flowchart of the steps involved in the model.
 }\label{fig:flowchart}
\end{figure*}

~\\
In summary, we incorporate the codes from the Sharma12 model into the L-Galaxies code, providing a new prescription for the radial distribution of hot halo gas in SAMs. In Fig. \ref{fig:flowchart}, we show a flowchart to briefly summarize the steps involved in the models described in this section. Here we briefly describe the steps that we follow:\\
(i) In each time step during the model calculation, we obtain the values of halo mass $M_{200}$, baryon fraction $f_{\rm b}$, concentration parameter $c_{200}$, dropout rate $f_{\rm d}$ and redshift $z$, and adopt them as inputs into the Sharma12 model.\\ 
(ii) We construct a set of concentric shells around each halo and calculate the radial profiles of hot gas density $\rho_{\rm hot}$ and gas temperature $T_{\rm hot}$, following the prescription from Sharma12.\\
(iii) Based on the new profiles of $\rho_{\rm hot}$ and $T_{\rm hot}$, we calculate the gas cooling, feedback, gas stripping and other baryonic processes of galaxy formation.\\
(iv) We save the model outputs, and calculate the X-ray luminosity and temperature based on the outputs of the hot gas profiles with Eq. \ref{eq:Lx} \& \ref{eq:Tx}.

\section{Model Results}

\subsection{The radial profiles of hot gas} \label{sec:profilesresult}

In Sharma12, the radial profiles in the model show consistency with observations of haloes with a given mass and concentration parameter. In this section, we compare the radial profiles of the hot gas from a large sample of galaxies and clusters in the L-Galaxies SAM to data from X-ray observations. This is a test on the validity of the new radial distribution model of the halo hot gas.

\begin{figure*}
 \includegraphics[angle=0,scale=0.44]{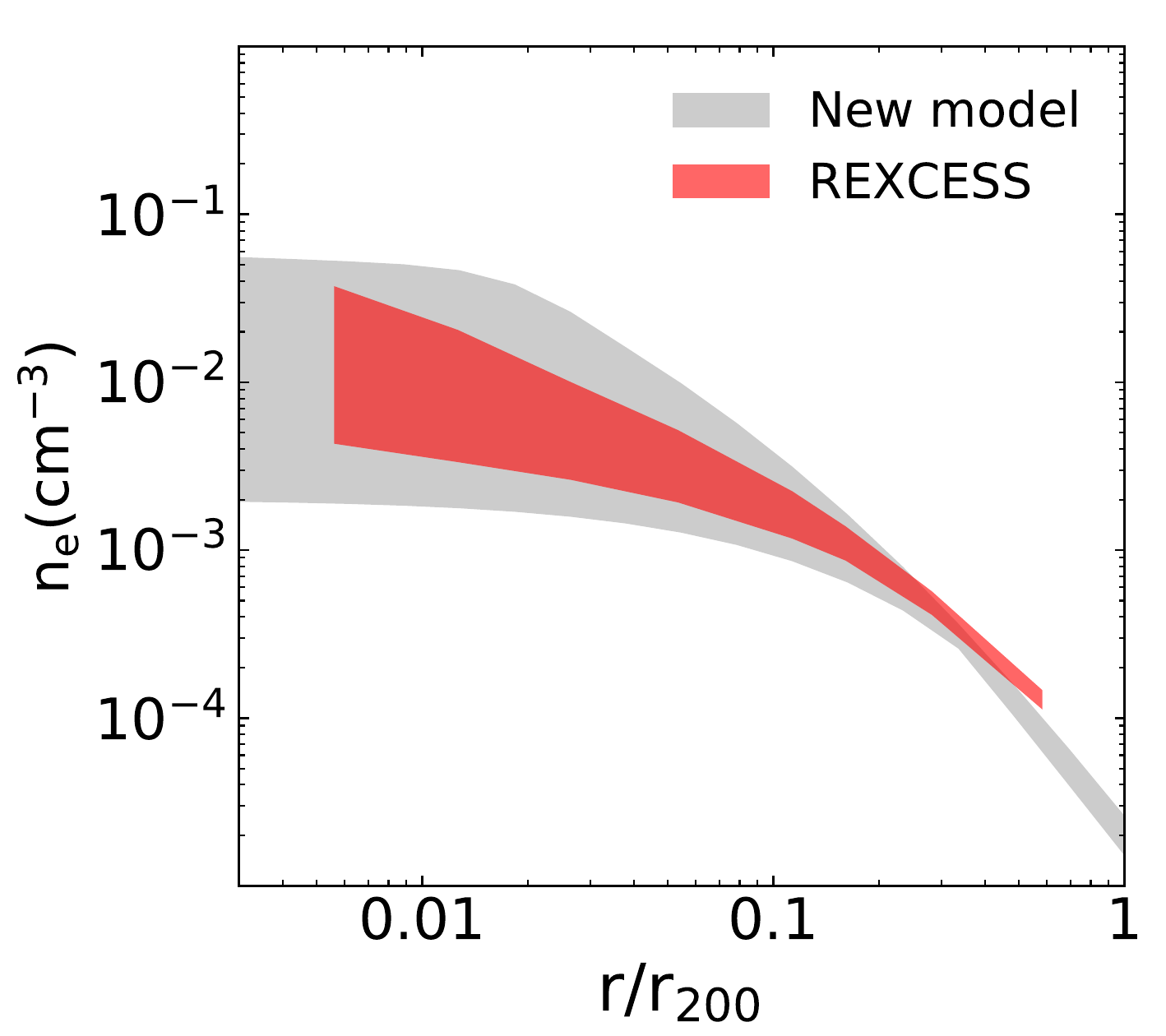}
 \includegraphics[angle=0,scale=0.4]{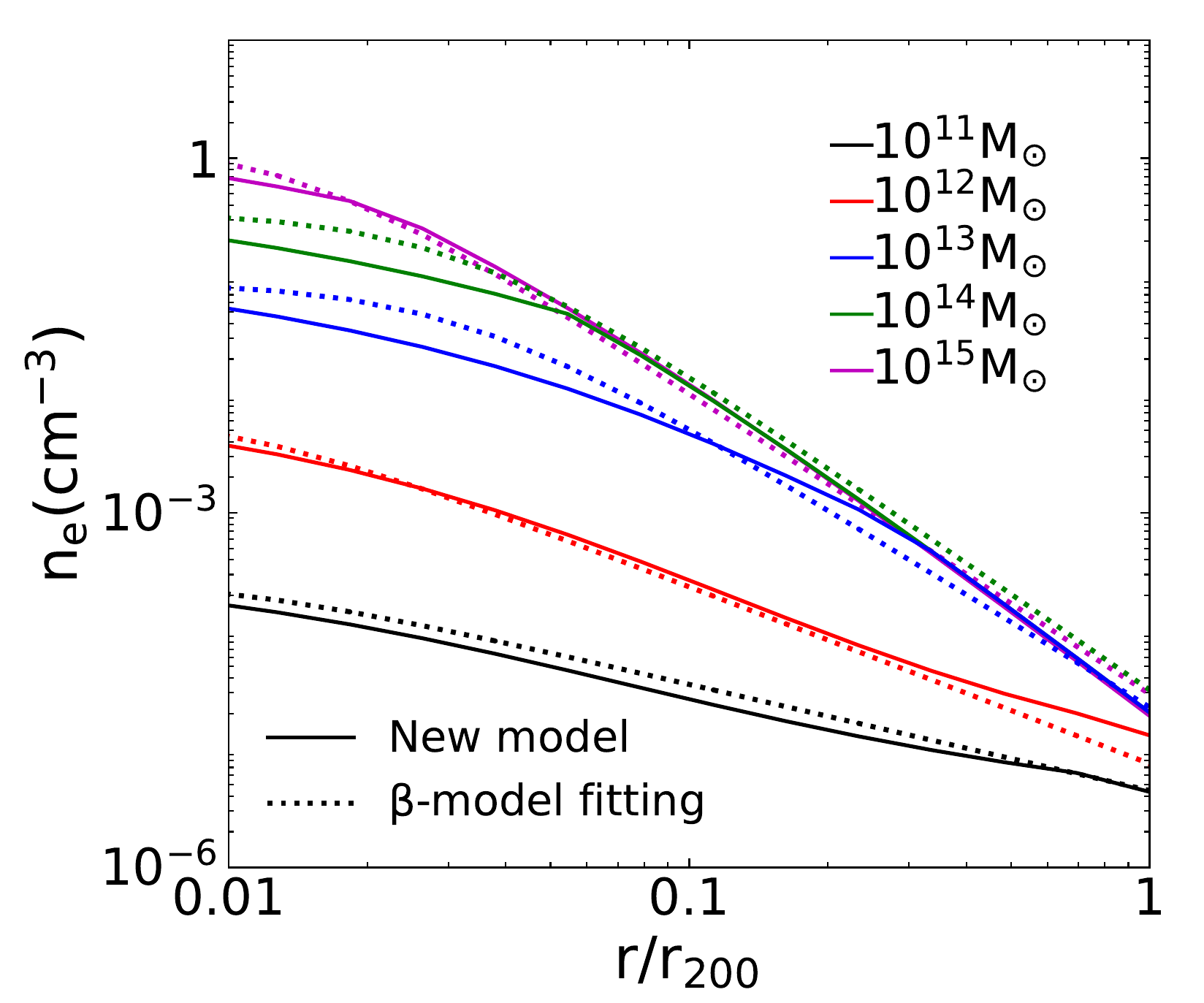}
 \caption
 {Left panel: The model results of electron density profiles compared with the data from REXCESS cluster sample (Croston et al. 2008). The red area shows the profiles from a representative sample of 31 nearby clusters by \emph{XMM-Newton} in $0.3-2.0$ keV band. The gray area represents the model results from massive clusters with large elliptical galaxies ($\Mvir>3\times10^{14}\ms$ and $c_{200}<7$) at $z=0$. In the observational data from Croston et al. (2008), the radii are in the unit of $r_{500}$, and we convert it to $r_{\rm 200}$ by $r_{\rm 200}=r_{500}c_{200}/c_{500}$ to compare with the model results.
 Right panel: The model results of electron density profiles (solid curves) compared with $\beta$ profiles (dotted curves). The curves in different colours represent the results from different mass haloes, and the model parameters here are the same as in Fig. \ref{fig:neprofiles}.} \label{fig:netempprofiles}
\end{figure*}

\begin{figure}
\centering
 \includegraphics[angle=0,scale=0.45]{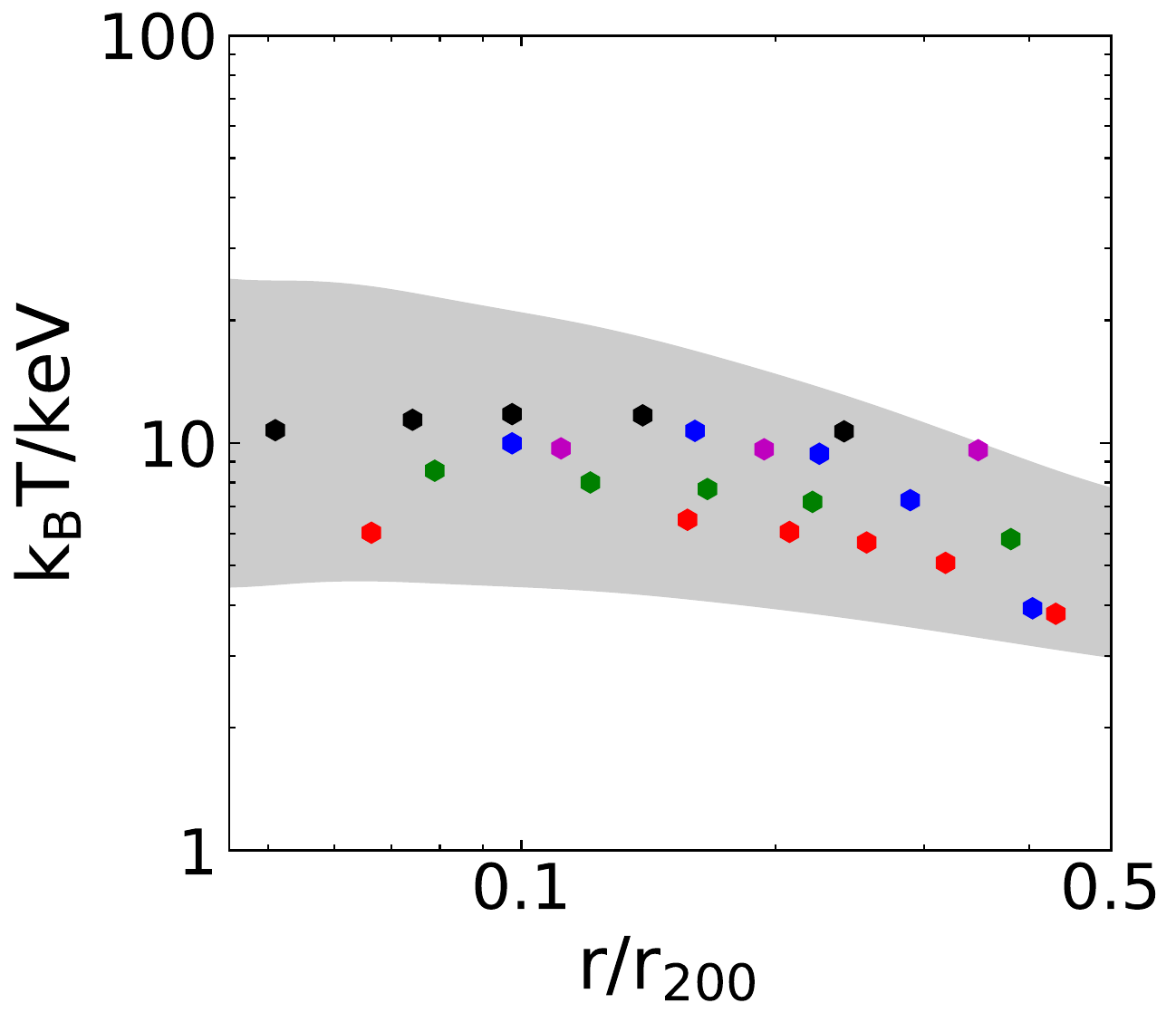}
 \caption{The new model results of hot gas temperature profiles compared with the observations. The gray area is from the same model sample as the left panel of Fig. \ref{fig:netempprofiles} ($\Mvir>3\times10^{14}\ms$ and $c_{200}<7$ at $z=0$). The data points in different colours are from 5 clusters observed by \emph{XMM-Newton} and \emph{Chandra} (Bartalucci et al. 2017). Similar to the left panel of Fig. \ref{fig:netempprofiles}, here we also convert the unit $r_{500}$ to $r_{\rm 200}$. 
}\label{fig:tprofile}
\end{figure}

In the left panel of Fig. \ref{fig:netempprofiles}, we compare the radial profiles of the hot gas electron density with the observational data from REXCESS (Croston et al. 2008), which observed 31 nearby clusters with the \emph{XMM-Newton} telescope. 
For comparison, we choose corresponding samples from massive clusters in our model at $z=0$. The density profiles become flat in the inner region ($r<0.1r_{\rm 200}$), which is approximately consistent with the trend from the observational data. According to Fig. \ref{fig:neprofiles}, this is expected since we adopt a core 
in the inner halo, \ie{}a ``cool-core cluster'' (Hudson et al. 2010), which depresses the density in the centre.

In the right panel of Fig. \ref{fig:netempprofiles}, we compare the electron density profiles of the model at different halo masses with the $\beta$ profiles. We adopt Eq. \ref{eq:betaprofile} to get the best-fitting parameters. The model curves here are the same as in the right panel of Fig. \ref{fig:neprofiles}. We can see that the $\beta$ model roughly fits the model profiles for haloes in quite a wide mass range. Since the $\beta$ profile is a widely used model that can fit many observations (e. g., Mirakhor et al. 2021, Bregman et al. 2018), the hot gas density profiles in our model can be 
easily compared with observations.

In Fig. \ref{fig:tprofile}, we show the hot gas temperature profiles from the new model compared with the observation from \emph{XMM-Newton} and \emph{Chandra} by Bartalucci et al. (2017). We can see that the gas temperature profiles from both the model and observations are flat or have a downturn in the innermost region of the halo ($r<0.1r_{\rm 200}$). This corresponds to the ``cool core'' in Fig. \ref{fig:coolmodel}. Similar results are also found by other works (\eg{}Vikhlinin et al. 2006). For the outer haloes at $r>0.5r_{\rm 200}$, the new model shows an upturn in temperature profiles in massive haloes ($\Mvir\gtrsim10^{13}\ms$) (see Fig. \ref{fig:tprofiles}), where the X-ray telescopes can hardly detect the emission from hot gas because of the low density. This is partly due to the imposed pressure boundary condition at $r_{\rm 200}$. 

When we compare $T_{\rm hot}$ with the viral temperature $T_{\rm 200}$ (see Fig. \ref{fig:tprofiles}), both the results from our new model and 
the observations show that $T_{\rm hot}$ is significantly higher than $T_{\rm 200}$ in inner haloes. Since $T_{\rm 200}$ is used as hot gas temperature in the isothermal model, we expect that higher $T_{\rm hot}$ will affect the processes of gas cooling and feedback in the new model. In the next part, we will discuss the effect of the new gas density and temperature profiles on the global X-ray properties in the model.

\subsection{The X-ray luminosity and temperature of hot gas} \label{sec:Lx}

The main purpose of this paper is to construct reasonable physical prescriptions of the hot gaseous haloes in SAMs. 
In this part, we will first describe how we calculate the X-ray emission from the halo hot gas, and then show some model results compared with the observations.

In previous L-Galaxies models, the bolometric X-ray luminosity is determined by a galactic cooling flow within the cooling radius (White \& Frenk 1991)
\begin{equation} \label{eq:cooling1991}
\LX=2.5\dot{m}_{\rm cool}v_{\rm 200}^2,
\end{equation}
in which the factor 2.5 represents a steady isobaric cooling flow. When the gross cooling rate is used in this equation, all the thermal energy of the cooling gas within the isothermal distribution is assumed to contribute to the X-ray emission. However, in the H15 model (and some other SAMs, \eg{}SAGE, Croton et al. 2016), because of the separation in the calculation of cooling and reheating, the X-ray luminosity calculated by Eq. \ref{eq:cooling1991} in this way can only be regarded as a strong upper limit.

We should also note that no X-ray luminosities are calculated for systems with strong AGN in Eq. \ref{eq:cooling1991} because cooling is shut down by the AGN feedback (Eq. \ref{eq:netcoolingratenew} in Sec. \ref{sec:AGN}). This makes like-for-like comparisons with observational data very difficult for the cooling flow described by Eq. \ref{eq:cooling1991}.

In this paper, we calculate the X-ray luminosity based on the local gas density of the hot gas based on the results of the density profiles, and the X-ray luminosity of the hot halo is
\begin{equation}\label{eq:Lx}
\LX=4\pi \int {{n_e}(r){n_i}(r)\Lambda \left( {T(r),Z} \right){r^2}dr} ,
\end{equation}
where the cooling function ${\Lambda} (T(r),Z)$ is from Sutherland \& Dopita (1993), and the local electron density $n_e(r)$, ion density $n_i(r)$, and gas temperature $T(r)$ can be obtained at different radius in each halo.

\begin{figure}
\centering
 \includegraphics[angle=0,scale=0.35]{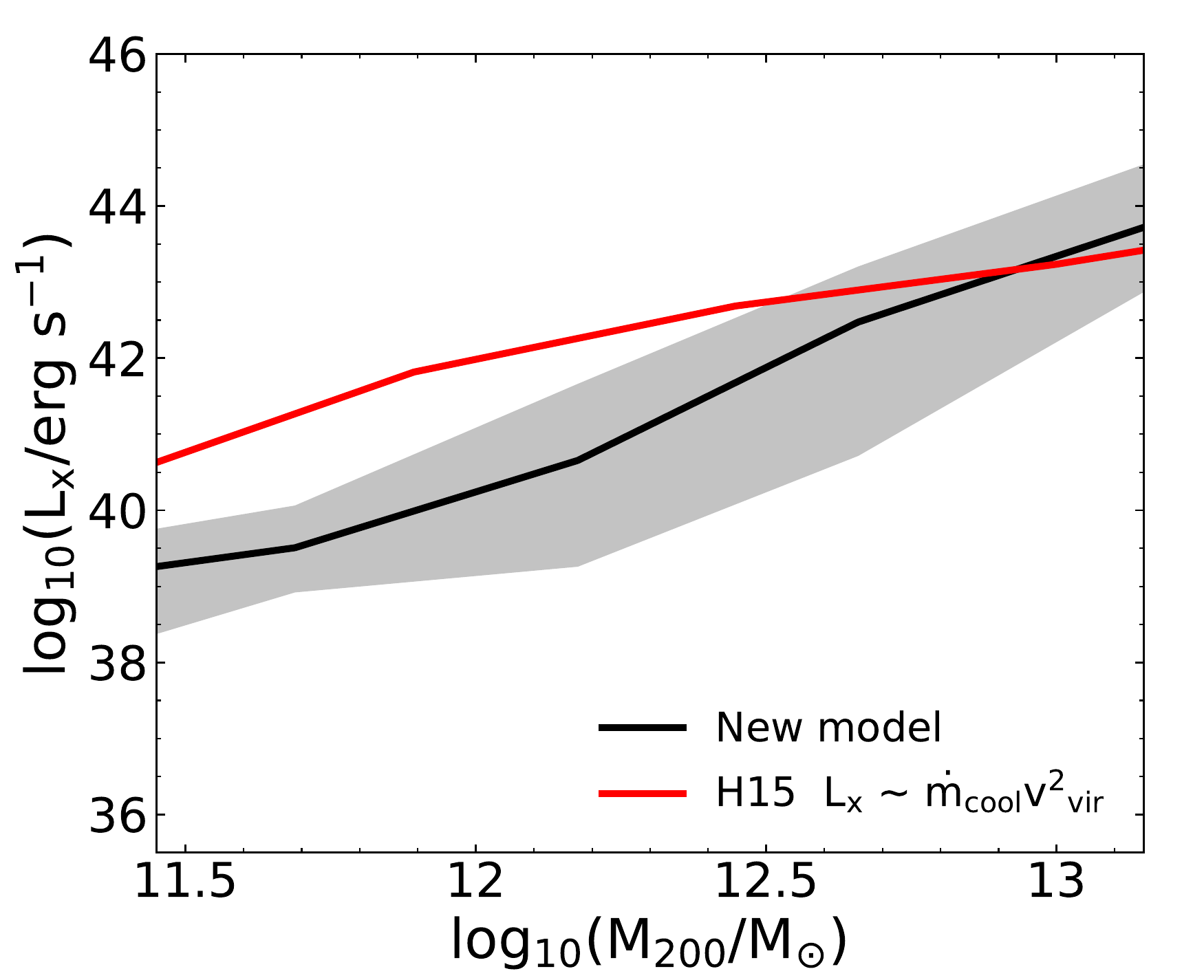}
 \caption{The $\LX$-$M_{\rm 200}$ relation from the model results at $z=0$. The solid curves represent mean values of the model results: red curve is the cooling luminosity from H15 model (Eq. \ref{eq:cooling1991}); black  curve is from the new model. The shaded area around the black curve represents $\pm1\sigma$ deviations for the samples of the new model.
 }\label{fig:Lxmvir}
\end{figure}

\begin{figure*}
\centering
 \includegraphics[angle=0,scale=0.45]{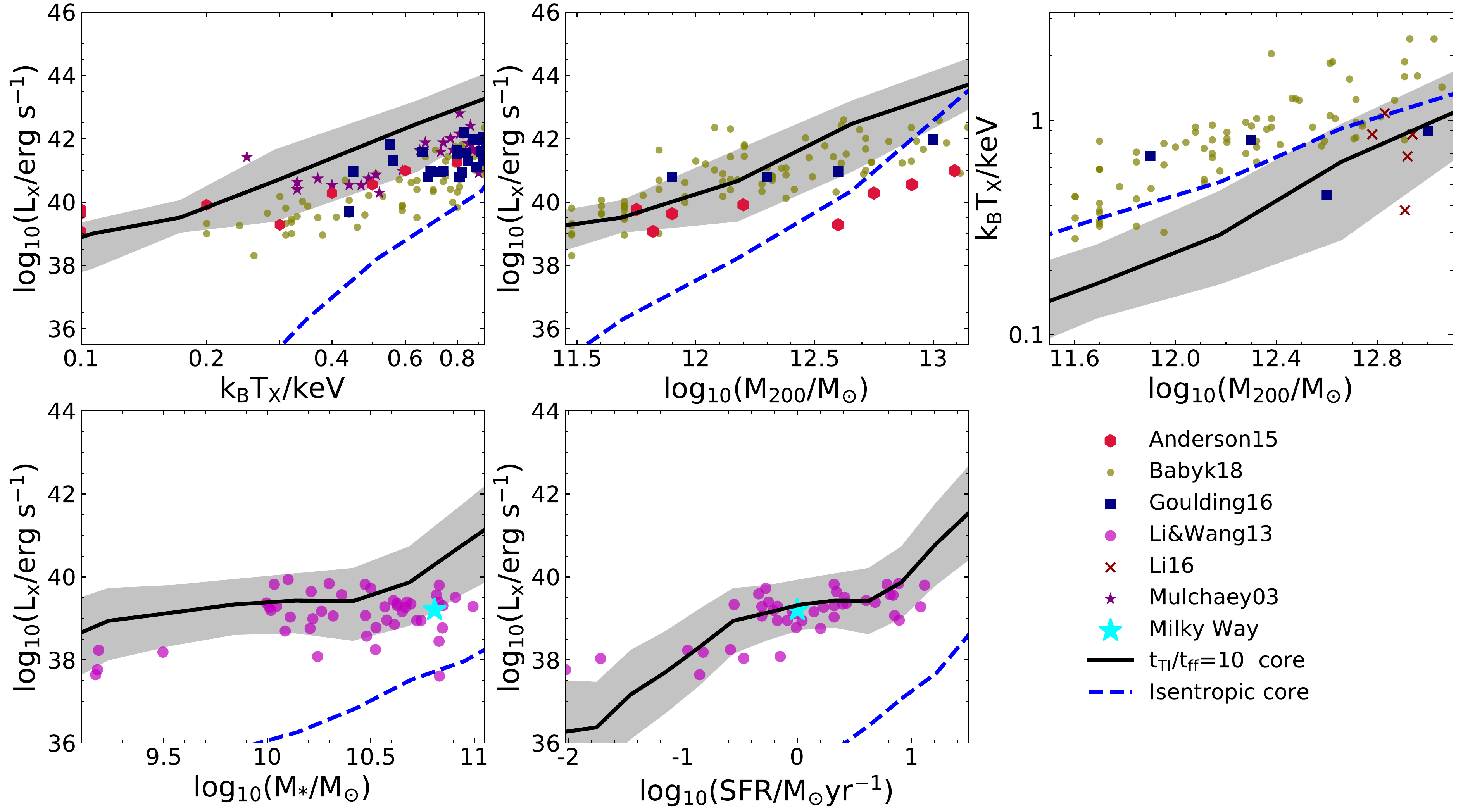}\\
 \caption{Top panels: The scaling relations of $\LX$, $\TX$ from the halo hot gas and the halo mass $\Mvir$. Bottom panels: $\LX$ from halo hot gas vs stellar mass and star formation rate.
 In each panel, the data points are from X-ray observations of galaxy clusters or groups. The model results are from the outputs at $z=0$, and the curves represent mean values of the model results: black solid curves are from the our new ``fiducial'' model, i. e.,  the cool core with $t_{\rm TI}/t_{\rm ff}=10$; the blue dashed curves are from the model results with an isentropic cool core. The shaded areas around the black curves represent $\pm1\sigma$ deviations for the model samples. The model results of $\LX$ and $\TX$ are from the halo regions within $0.3r_{\rm 200}$. For the bottom two panels, the model results are from central galaxies with $\rm{SFR}/M_*>10^{-11}\rm{yr}^{-1}$ at z=0, and we also include the values from Milky Way for comparison (Snowden et al. 1997).
 }\label{fig:Lxhalorelation}
\end{figure*}

In Fig. \ref{fig:Lxmvir}, we show comparison of the X-ray luminosity between the isothermal sphere in H15 and the new model for the results at $z=0$. As desired, $\LX$ from our new model is lower than the cooling luminosity from isothermal sphere, especially for low mass haloes ($M_{\rm 200}\lesssim10^{12.5}\ms$), which is one of the motivations to update the model in this paper. The main cause of the decrease of $\LX$ is the reduction of hot gas density, especially in the inner haloes. In Fig \ref{fig:neprofiles}, we can see that the gas density in inner haloes show a flatter ``core'' in Sharma12 model rather than a ``cusp'' in the isothermal profile. Since $\LX\sim n_{e}^2$ is mainly contributed by hot gas emission from the inner halo, the flatter gas profile in inner core gives much lower $\LX$ for the whole halo. The new model can give accurate X-ray luminosity compared to the isothermal sphere (see Fig. \ref{fig:Lxhalorelation}), and we will then only consider $\LX$ from the new model in the following paper.

Another important property usually adopted in X-ray observations is the X-ray luminosity-weighted temperature of a halo, $\TX$. Similar to Eq. \ref{eq:Lx}, we get $\TX$ in the model from the density and temperature profiles of the hot gas,
\begin{equation}\label{eq:Tx}
{T_{\rm{X}}} = L_{\rm{X}}^{ - 1}4\pi\int {T\left( r \right)n_e(r)n_i(r)\Lambda \left( {T\left( r \right),Z} \right) {r^2}dr},
\end{equation}
where $L_{\rm{X}}$ is the total X-ray luminosity from halo hot gas in Eq. \ref{eq:Lx}.

In the top three panels of Fig. \ref{fig:Lxhalorelation}, we show the X-ray luminosity $\LX$ and temperature $\TX$ from the halo hot gas as a function of the halo mass $\Mvir$, and also show the scaling relation of $\LX$ vs $\TX$. The data points are from various X-ray observations: Goulding et al. (2016) and Babyk et al. (2018) are from \emph{Chandra}; Mulchaey et al. (2003) and Anderson et al. (2015) are from \emph{ROSAT}; Li et al. (2016) is from \emph{XMM-Newton}. For the model results, we show the black curves from our new model from the central galaxies at $z=0$. In each panel, we calculate $\LX$ and luminosity-weighted temperature $\TX$ by Eq. \ref{eq:Lx} \& \ref{eq:Tx}. To be consistent with the observations, we integrate the $\LX$ and $\TX$ from the model profiles in the region of $r<0.3r_{\rm 200}$, since most of the X-ray data extend only to the inner halo around $0.3r_{\rm 200}$ or $r_{500}$. We also show the relation of $\LX$ as a function of stellar mass $M_*$ and star formation rate SFR in the bottom two panels of Fig. \ref{fig:Lxhalorelation} compared to the \emph{Chandra} data points from Li \& Wang (2013). Latter are mainly the results from the halo hot gas of star-forming disk galaxies, and we choose the central galaxies with $\rm{SFR}/M_{*}>10^{-11}\rm{yr}^{-1}$ from the model samples at $z=0$.

In Sec. \ref{sec:hotprofilemodel}, we mention that we adopt the gas profiles with ``$t_{\rm TI}/t_{\rm ff}=10$ core'' other than isentropic core in this paper. For comparison, we also plot the results of the scaling relations with an isentropic core in Fig. \ref{fig:Lxhalorelation}. The new model with ``$t_{\rm TI}/t_{\rm ff}=10$ core'' gives better results for the X-ray luminosity, especially at the low mass end. For the isentropic model, the slope is too steep for the $\LX-\Mvir$ and $\LX-\TX$ relations, 
which leads to an underestimate of $\LX$ for low mass haloes/galaxies. The main reason is that the model with isentropic core gives too low and flat gas density profiles for haloes with $\Mvir\lesssim 10^{12}\ms$ (see the left panel of Fig. \ref{fig:neprofiles}). Thus, we adopt the cool core with ``$t_{\rm TI}/t_{\rm ff}=10$ core'' as the ``fiducial'' model in this paper.

In Fig. \ref{fig:Lxhalorelation}, the new model can quite satisfactorily reproduce the data from various observations. The slopes of the scaling relations are $\LX\sim\Mvir^{2.5}$, $\LX\sim\TX^{4.5}$, and $\Mvir\sim\TX^{2.0}$, which is very similar to the slopes from \emph{Chandra} by Babyk et al. (2018), \ie{}$\LX\propto\Mvir^{2.8\pm0.3}$, $\LX\propto\TX^{4.5\pm0.2}$, and $\Mvir\propto\TX^{2.4\pm0.2}$. For haloes in self-similar collapse model with pure bremsstrahlung emission, $\LX\propto\Mvir^{4/3}$ and $\LX\propto\TX^{2}$ (Pratt et al. 2009). The Sharma12 model predicts $\LX\sim\TX^{3}$ for massive haloes, because of lower gas densities in smaller haloes (see Section 3.1 in Sharma12). When we consider a wider halo mass range, the slope of $\LX-\TX$ relation becomes even steeper because $T_{\rm gas}$ tends to be a lot higher than $T_{\rm 200}$ in small haloes.

For the gas temperature in Fig. \ref{fig:Lxhalorelation}, the $\TX$ predicted by the new model is about 3 to 10 times higher than $T_{\rm 200}$. We turn to Fig. \ref{fig:tprofiles} again and can see that the main cause for a higher $\TX$ than $T_{\rm 200}$ is $T_{\rm gas}>T_{\rm 200}$ in the inner halo, which corresponds to the cool core region undergoing thermal instability and gas infall. For small haloes in fast mode cooling (\ie{}haloes with $r_{\rm cool}=r_{\rm 200}$), the ratio of $\TX/T_{\rm 200}$ tends to be larger than haloes in slow mode cooling ($\Mvir\gtrsim10^{13}\ms$), and $\TX/T_{\rm 200}\propto\Mvir^{1/2}/\Mvir^{2/3}\propto\Mvir^{-1/6}$ from slope of the $\TX-\Mvir$ relation mentioned above.

\section{The baryon budget and the missing baryon problem}

In this section, we further study the baryon budget in different components of model galaxies, and then explore how the missing baryon problem manifests in our new model.

As mentioned in Sec. \ref{sec:model}, the baryonic components in L-Galaxies models can be divided into three sets of components: the baryons in galaxies (star, ISM cold gas, black hole), baryons in haloes (ionized hot gas), and baryons out of haloes (the external reservoir ejected by supernovae, which corresponds to the ``dropout mass'' in the Sharma12 model).

\begin{figure}
\centering
 \includegraphics[angle=0,scale=0.4]{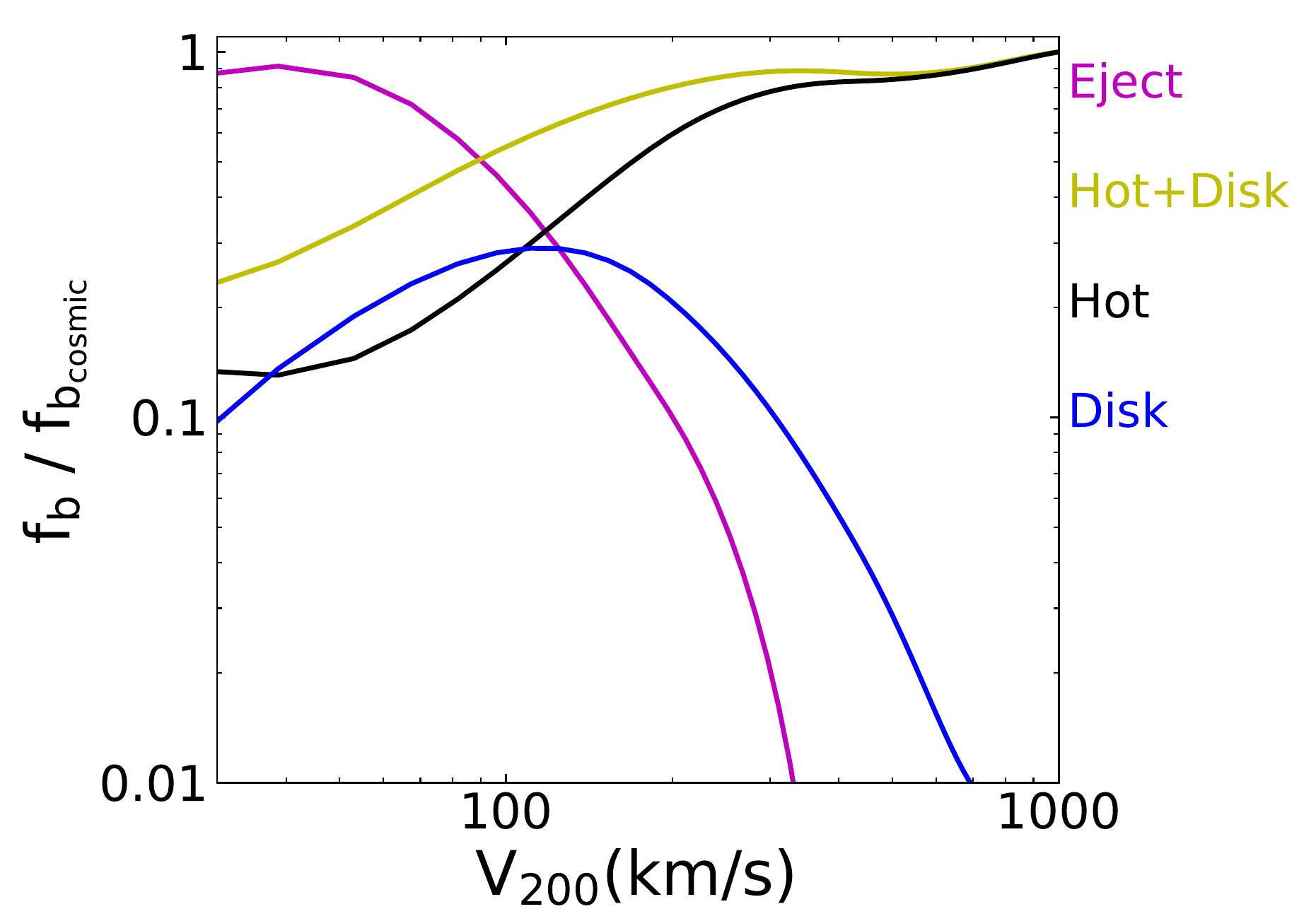}
 \caption{The baryon fraction of different components from model results at $z=0$. The x axis is the virial velocity of the central halo, and the y axis is the baryon fraction normalized to cosmic baryon fraction. All the solid curves represent the mean values from the model results. The blue curve represents the baryon fraction of disk components, including the baryon mass in stars, cold gas and black holes. The black curve presents the baryon fraction of halo hot gas. The yellow curve represents the sum of the disk and halo components. The purple curve represents the external reservoir beyond haloes. 
 }\label{fig:fbaryon}
\end{figure}

In Fig. \ref{fig:fbaryon}, we show the relation between the baryon fraction
\begin{equation} \label{eq:fbcomp}
f_{\rm b, comp}=\frac{m_{\rm b, comp}}{m_{\rm halo}}
\end{equation}
and the virial velocity of haloes, where
the subscript ``comp'' can be ``disk'' (the sum of star, cold gas and black hole), ``hot'' (hot gas in halo) or ``eject'' (ejected reservoir).

For haloes with $v_{\rm 200}\gtrsim100$~km~s$^{-1}$, a large proportion of baryonic matter exists as hot gas in the halo, and $f_{\rm b, disk}$ becomes smaller in more massive haloes (blue curve) mainly due to supernova feedback following merger-induced starbursts, which ejects a large fraction of baryons out of discs and suppresses the subsequent gas cooling and infall (along with any AGN feedback). The blue curve in Fig. \ref{fig:fbaryon} is consistent with the observed baryon-to-halo mass relation (Papastergis et al. 2012; baryon refers to star+atomic gas in that paper) and the stellar-to-halo mass relation (\eg{}Girelli et al. 2020) in massive haloes.

In small haloes ($v_{\rm 200}\lesssim100$~km~s$^{-1}$), a large proportion of baryonic matter consists of star and cold gas in the galaxy disk, and the proportion in halo hot gas $f_{\rm b, hot}$ is quite low (smaller than $20\%$ of the cosmic baryon fraction). The timescale for gas cooling and infall is very short in these small haloes. Most are in the ``fast-mode cooling'' regime (see $r_{\rm cool}$ vs $r_{\rm 200}$ in Fig. \ref{fig:rcoolcompare}) and the hot gas is accreted onto the disk in a short time. On the other hand, supernova feedback can easily eject the baryons out of the halo because of the shallow potential wells of small haloes. Thus, a large proportion of the baryonic matter tends to be unbounded from the small haloes (purple curve in Fig. \ref{fig:fbaryon}). The unbounded reservoir consists of ionized gas with very low density and a high temperature, which corresponds to the cosmic web or filaments in diffuse IGM and WHIM (see the results from Illustris-TNG by Martizzi et al. 2019). Because of the low density, the hard-to-detect ionized gas in $f_{\rm b, eject}$ should be one of the main components of missing baryons (Shull et al. 2012), which can be observed by stacking of X-ray images or via the Sunyaev-Zeldovich effect (Wang et al. 2019).

\begin{figure}
\centering
 \includegraphics[angle=0,scale=0.37]{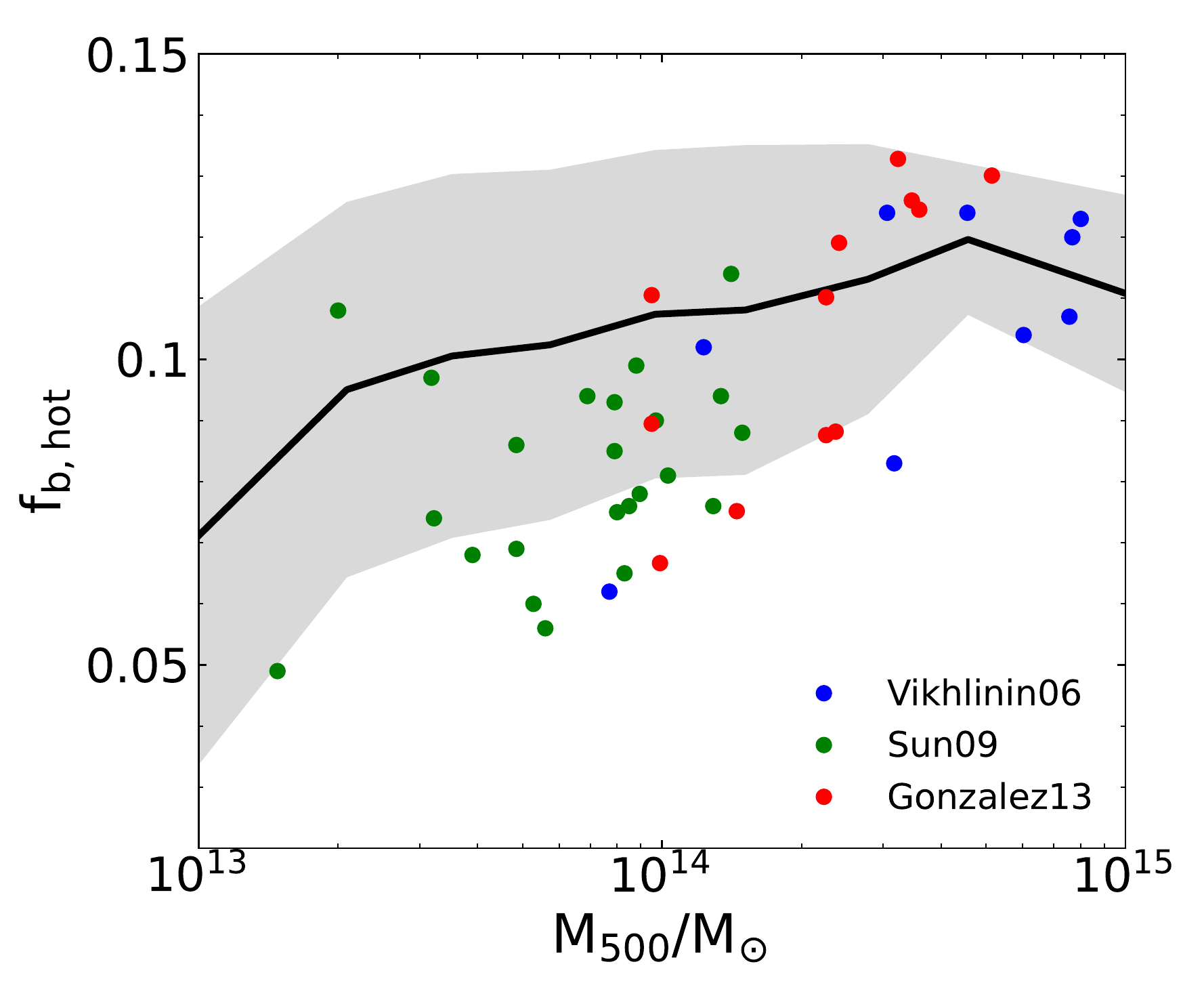}
 \caption{The model results of baryon fraction in hot gas components at $z=0$ compared with X-ray data. Here, $f_{\rm b, hot}$  is defined as the ratio of hot gas mass to the halo mass inside $r_{500}$, and the data points are from X-ray observations of galaxy groups and clusters. For model results, we convert the halo mass $m_{\rm 200}$ to $m_{500}$ in the X-scale and only include the mass of hot gas with $T_{\rm gas}$ above 0.5~keV to be consistent with the observational samples. The black curve with shaded area represent the mean values with $\pm 1\sigma$ deviations of the model results.
 }\label{fig:fbhot}
\end{figure}

Then we focus on the hot gas components in haloes (the black curve in Fig. \ref{fig:fbaryon}). To test the model results, in Fig. \ref{fig:fbhot} we compare the baryon fraction in halo hot gas with X-ray observations. The data points are from X-ray observations of hot gas components in galaxy groups and clusters: Gonzalez et al. (2013) is from \emph{XMM-Newton}; Vikhlinin et al. (2006) and Sun et al. (2009) are from \emph{Chandra}. To be consistent with the observations, the baryon fraction in hot gas $f_{\rm b, hot}$ is defined as the ratio of hot gas mass to halo mass inside $r_{500}$, and we calculate $m_{500}$ with $m_{500}=0.72m_{\rm 200}$ (Pierpaoli et al. 2003). For possible detection in the X-ray observations, we only include the mass of hot gas with $T_{\rm gas}>0.5$~keV. In Fig. \ref{fig:fbhot}, we can see that the model results roughly match the observations, which shows the validity of our new model and the validity of baryonic processes in current SAMs. For example, similar results were also found by Yates et al. (2017), when modifying the infall prescription in an earlier version of L-Galaxies and assuming a mass-dependent hot gas density profile fit to observations.

\begin{figure}
\centering
 \includegraphics[angle=0,scale=0.35]{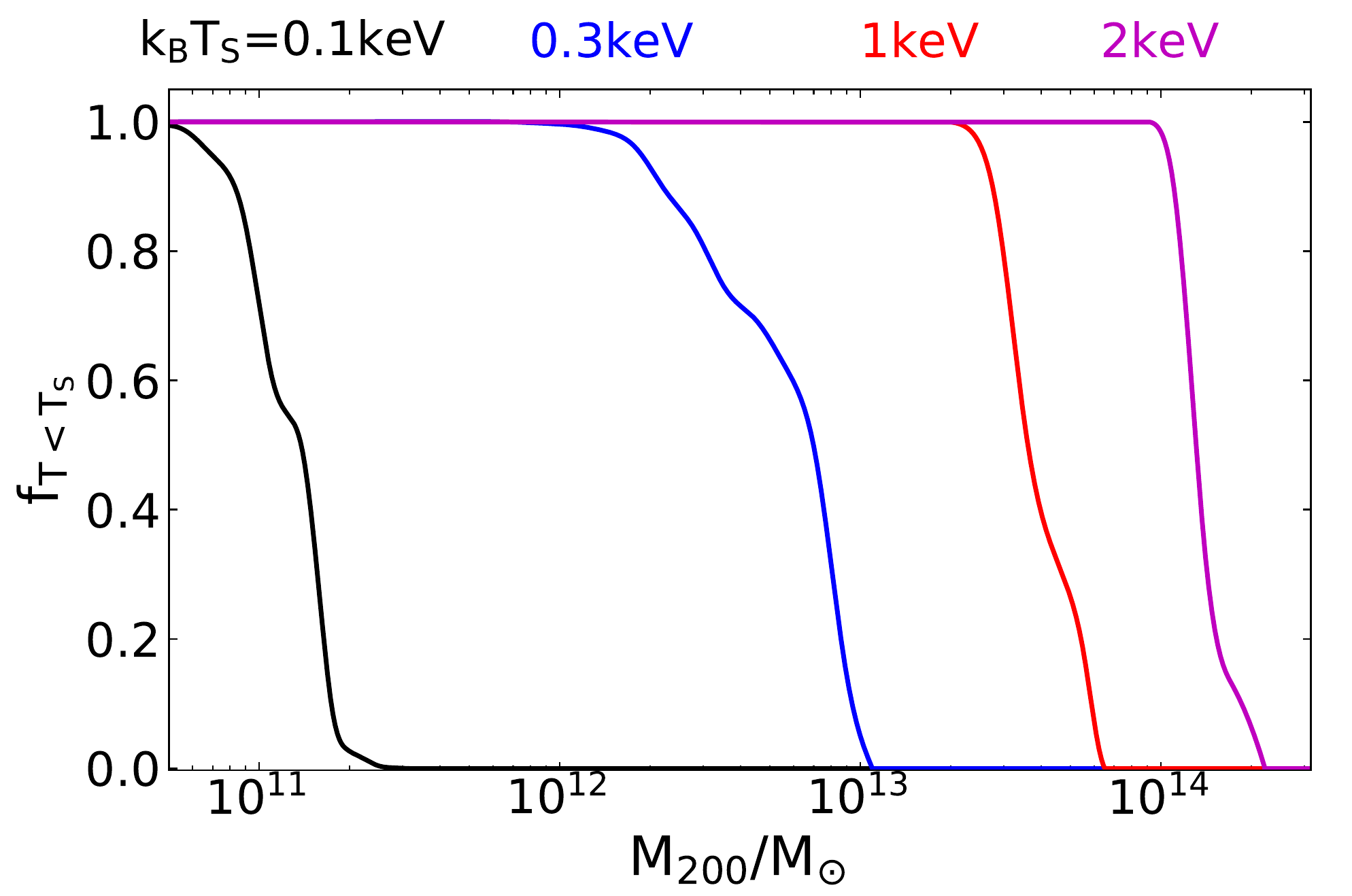}
 \caption{The mass fraction of the halo hot gas below the given temperature ($T_{\rm S}$) vs halo mass from the model results at $z=0$. The four curves represent the fractions of the halo gas lower than 0.1, 0.3, 1.0 and 2.0~keV.
 }\label{fig:fhotgas}
\end{figure}

Based on the hot gas temperature and density profiles in our new model, we plot the mass fraction of the halo hot gas below a given temperatures $T_{\rm S}$ as a function of halo mass, and we choose some common limits in the X-ray band as the values of $T_{\rm S}$. As shown in Fig. \ref{fig:fhotgas}, a large fraction of the hot gas in massive haloes ($\Mvir>10^{14}\ms$) has a temperature above 2.0~keV, which should be relatively easy to detect by most X-ray telescopes. For smaller haloes, \eg{}those around the Milky Way halo mass ($3\times10^{12}\ms$), the temperature of most halo hot gas is below 0.3~keV, which is close to the detection limit in soft X-ray band (the low energy band is 0.3-10~keV in eROSITA and 0.3-10~keV in Athena, Kaastra et al. 2013). 
For haloes with $\Mvir\sim10^{11}\ms$ or $V_{\rm 200}\lesssim100~\rm{km~s}^{-1}$, a large fraction of the gas has temperature around 0.1~keV or even lower, which corresponds to the low temperature WHIM or CGM. The soft X-ray emission from these gas components is difficult to detect by current facilities because of the partial ionization state, moderate temperature, and low density (Paerels et al. 2008). None the less, various simulations suggest that this component should contain a significant amount of the total metals in the Universe (Yates et al. 2021b). Tracing metal emission or absorption lines (e. g., OVII and OVIII) in soft X-ray band (Shull et al. 2012, Nicastro et al. 2018) or the absorption lines in the UV band (Stern et al. 2016, Zahedy et al. 2019) should be the main method to detect these gas components.

According to the discussion above, the ionized gas in the unbounded reservoir out of halo and low temperature intergalactic gas bound in low mass haloes should be the main components of the ``missing baryons''. Combining Fig. \ref{fig:fbaryon} \& \ref{fig:fhotgas}, the fraction of these hard-to-detect baryons tend to be higher around low-mass haloes, which is consistent with baryon-to-halo mass relation (Dai et al. 2010, McGaugh et al. 2010), and the missing baryons mainly exist in and around small haloes (McGaugh 2008). In our subsequent work, we will give more detailed predictions for the observations of these hot gas components and produce mock X-ray surveys based on the model outputs.

\section{Summary and conclusions}

In this paper, we extend the L-Galaxies SAM to include a new model for the radial distribution of hot gas. This model considers local instabilities and thermal equilibrium in the halo. This replaces the simple isothermal sphere applied in previous models. The main motivation of this paper is to offer the results of X-ray properties and study the physical mechanisms related to halo hot gas and baryon budgets in the framework of SAMs.

The main changes made to the L-Galaxies SAM are as follows:

\noindent (i) Each hot gaseous halo is described by a series of concentric shells, so that the radial profiles of gas density and temperature in each halo are obtainable, and the processes affecting the hot halo gas can be based on the new gas density profiles.

\noindent (ii) We adopt the model from Sharma12 instead of the isothermal sphere to describe hot gas distribution. In this model, the local thermal instability time-scale $\tti$ and free-fall time-scale $\tff$ of the gas must satisfy $\tti/\tff\gtrsim10$ everywhere. According to the ratio  $\tti/\tff$ at each radius, the hot gas halo can be divided into two regimes: a cool core with a flatter profile in the central part, and the stable region in the outer halo.

\noindent (iii) We update the prescriptions for gas cooling and infall, which are now designed to mimic the cooling of gas through the formation of filaments and blobs in the cool core due to thermal instabilities. The ``fast-mode cooling'' regime corresponds to a scenario where $r_{\rm cool}=r_{\rm 200}$, and the ``slow-mode cooling'' regime corresponds to the scenario where $r_{\rm cool}<r_{\rm 200}$. In both modes, the hot gas in the cool core falls into the central galaxy in one free-fall time scale.

\noindent (iv) We update the prescriptions for supernova feedback, radio-mode AGN feedback, and gas stripping in satellites based on the new gas temperature and density profiles. In the supernova reheating and radio-mode AGN accretion processes, we assume the ejected gas to have the same specific thermal energy as the halo hot gas instead of the virial temperature used in previous models.

\noindent (v) We adopt a simple model to describe the atomic and molecular gas transition in the ISM, and the star formation rate is related to the $\h2$ mass, according to recent observations.

\noindent (vi) We also tune the model parameters to fit various observations.

Based on the new model results and physical prescriptions, the main conclusions of this paper are:

\noindent (i) The new model returns 
a much better match to X-ray observations compared with the previous model. The main reason for this is flatter cores in the inner halo, rather than the ``cusps'' present in the isothermal sphere approximation.

\noindent (ii) The temperature of the hot gas is higher than the virial temperature in most haloes, which is mainly caused by the high density gas undergoing thermal instability and infall in the cool core region. A higher ratio of $\TX/T_{\rm 200}$ in smaller haloes and a lower density in the cool core leads to a steeper slope in the $\LX-\TX$ relation.

\noindent (iii) Our model suggests that the ionized gas in the unbounded reservoir out of halo potential and low temperature intergalactic gas bounded in low mass haloes should be the main components of the ``missing baryons''. The fraction of hard-to-detect baryons tend to be higher in lower mass haloes.

In summary, the models in this paper provide 
a faithful description of the ionized halo hot gas in SAMs. Based on the model outputs, we can make predictions for various observations of the hot gas surrounding galaxies. In particular, by taking advantage of the large box size of the L-Galaxies SAM, we plan to construct mock observations for large scale surveys of the baryons in soft X-ray band, which is proposed by some X-ray telescopes in the future, \eg{}the study of the WHIM by Athena X-ray Observatory (Kaastra et al. 2013) and Hot Universe Baryon Surveyor (HUBS Satellite, Cui et al. 2020) .

Finally, we should caution that the models to describe the hot gas in this paper are simple in their construction. We assume the gas is in a spherical distribution and only consider one-dimensional profiles of the hot gas halo. We do not include detailed structures in SAMs, like filaments, knots or the cosmic web, even though some hydrodynamic simulations propose that WHIM or missing baryons are mainly located in these structures (\eg{}Illustris-TNG by Martizzi et al. 2019, BAHAMAS by McCarthy et al. 2017).
On the other hand, some works suggest that AGN feedback affects the X-ray luminosity of haloes to some extent (\eg{}Puchwein et al. 2008, Gaspari et al. 2014, and the cool core cycles explored by Prasad et al. 2015), while our current model does not include 
the direct influence of AGN feedback on the X-ray emission. 
It should be interesting to explore the direct effect of AGN on $\LX$ from halo hot gas in the future.

\section*{Acknowledgments}

The authors thank the helpful suggestions by the anonymous referee. WXZ and JF acknowledge the support from Shanghai Committee of Science and Technology grant No.19ZR1466700, the fund for key programs of Shanghai Astronomical Observatory E195121009, the Youth innovation Promotion Association CAS, and the National SKA Program of China No. 2020SKA0110102. JF thanks the invitation to visit Max-Planck Institute for Astrophysics to meet and start the cooperation with PS. PS acknowledges a Swarnajayanti Fellowship from the Department of Science and Technology, India (DST/SJF/PSA-03/2016-17), and a Humboldt fellowship for supporting his sabbatical stay at MPA Garching. We thank Dr. Luo Yu in Purple Mountain Observatory and Prof. Guinevere Kauffmann in Max-Planck Institute for Astrophysics for the helpful discussions and suggestions. 

\section*{DATA AVAILABILITY}

The data and codes in this paper will be shared on reasonable request to the corresponding author.

\def\apj{ApJ}
\def\apjl{ApJL}
\def\apjs{ApJS}
\def\aj{AJ}
\def\aap{A\&A}
\def\araa{ARA\&A}
\def\aapss{A\&AS}
\def\mnras{MNRAS}
\def\nature{Nature}
\def\apss{Ap\&SS}
\def\pasp{PASP}

{}

\end{document}